\documentclass[11pt,a4paper]{article}
\usepackage{amsmath,amssymb,amsfonts}
\usepackage{slashed}
\usepackage{geometry}
\usepackage{graphicx}
\graphicspath{{outputs/}{./}}
\usepackage{booktabs}
\usepackage[font=small,labelfont=bf]{caption}
\usepackage{cite}
\usepackage[colorlinks=true,citecolor=blue,linkcolor=blue,urlcolor=blue]{hyperref}
\usepackage{cleveref}
\usepackage{xcolor}
\usepackage[left]{lineno}
\numberwithin{equation}{section}

\newcommand{\TeV}{\text{TeV}}
\newcommand{\GeV}{\text{GeV}}
\newcommand{\fb}{\text{fb}}

\newcommand{\arxiv}[1]{\href{https://arxiv.org/abs/#1}{arXiv:#1}}
\newcommand{\doi}[1]{\href{https://doi.org/#1}{doi:#1}}

\title{Constraints on Vector-Like Top Dipole Interactions from
Top-Associated Photon Measurements at the LHC}
\author{M. Sahraei, Y. Hosseini, M. Mohammadi Najafabadi \\
School of Particles and Accelerators, Institute for Research in Fundamental Sciences (IPM),\\
P.O. Box 19395-5531, Tehran, Iran\\
{\small\texttt{(msahraei, yhosseini, mojtaba)@ipm.ir}}}
\date{}

\begin{document}
\maketitle

\begin{abstract}
Precision measurements of top-quark production with energetic photons provide
a complementary probe of vector-like quarks with nonstandard radiative
decays. We reinterpret the CMS differential $t\bar t\gamma$ measurement and
the ATLAS fiducial $t\bar t\gamma\gamma$ cross section to constrain the
electromagnetic and chromomagnetic transition interactions of a charge-$2/3$
vector-like top partner, $T$. We consider effective interactions written
in terms of physical fields after electroweak symmetry breaking, retaining
only the decays $T\to t\gamma$ and $T\to tg$. The CMS signal includes
both the mixed $(t\gamma)(tg)$ topology and double-radiative
$t\bar t\gamma\gamma$ events entering the exactly-one-photon fiducial
region, while the ATLAS rate is interpreted using a signal-specific
detector-response transfer factor. Signal uncertainties are included, and
the interpretation is restricted point by point to the domain satisfying
the EFT and narrow-width requirements. Denoting the electromagnetic and
chromomagnetic transition coefficients by $c_{t\gamma}$ and $c_{tg}$,
respectively, we find that, for
$c_{tg}=1.0\times10^{-5}~\GeV^{-1}$ and $m_T=500~\GeV$, the
observed Run~2 exclusions at 95\% confidence level extend down to
$c_{t\gamma}\simeq4.7\times10^{-7}~\GeV^{-1}$ for CMS and
$2.3\times10^{-6}~\GeV^{-1}$ for ATLAS within the valid domain.
At $m_T=1000~\GeV$, the corresponding values are
$2.1\times10^{-6}$ and $8.5\times10^{-6}~\GeV^{-1}$.
Projections to the HL-LHC indicate improved sensitivity to these
radiative transition interactions. These results establish top-associated
photon measurements as complementary probes of radiative top-partner
interactions beyond searches targeting conventional decay modes.
\end{abstract}

\section{Introduction}
\label{sec:intro}

Vector-like quarks (VLQs) are predicted in many extensions of the
Standard Model (SM), including composite-Higgs models
\cite{Kaplan:1983sm,Contino:2003ve,Vignaroli:2012sf}, theories with extra
dimensions
\cite{Randall:1999ee,Antoniadis:1990ew,Arkani-Hamed:1998jmv}, and
little-Higgs constructions \cite{Schmaltz:2005ky}. Recent reviews
summarize their broader phenomenology and current collider status
\cite{Banerjee:2024wvs,Banerjee:2022xmu}. In contrast to the chiral
fermions of the SM, the left- and right-handed components of a
vector-like fermion transform identically under the gauge group, allowing
a gauge-invariant Dirac mass before electroweak symmetry breaking. Among
the possible VLQ states, a heavy top partner $T$ with electric charge
$+2/3$ is particularly well motivated by its possible connection to the
dynamics that stabilize the Higgs sector and address the hierarchy
problem. Related vector-like matter can also occur in supersymmetric
extensions \cite{Martin:2009bg}; Ref.~\cite{Martin:1997ns} provides a
broader review of supersymmetric constructions. Vector-like fermions can
also modify the renormalization-group evolution of the Higgs quartic
coupling and thereby affect the stability of the electroweak vacuum
\cite{Xiao:2014kba}. \\
The ATLAS and CMS Collaborations have performed extensive searches for
single and pair production of vector-like $T$ quarks in the conventional
electroweak decay modes $T\to Ht$, $T\to Zt$, and $T\to Wb$
\cite{ATLAS:2014vpn,ATLAS:2016scx,ATLAS:2018tnt,CMS:2016edj,
CMS:2016jce,CMS:2017voh,ATLAS:2018ziw,ATLAS:2025bzt,
ATLAS:2018cye,CMS:2017ynm,CMS:2022yxp}. These searches span Run~1 and
Run~2 data and are interpreted under specified assumptions about the
electroweak representation, production mechanism, total width, and
branching fractions of the heavy quark. Pair-production searches also
provide interpretations for different mixtures of the $Wb$, $Zt$, and
$Ht$ branching fractions \cite{CMS:2022yxp}. Representative analyses
have reported the following observed 95\% confidence level (CL) lower limits
under the assumption of a 100\% branching fraction into the indicated
mode \cite{ATLAS:2018tnt,ATLAS:2018cye,CMS:2017ynm}:
\begin{align}
m_T &> 1430~\GeV &&\text{for }T\to Ht,
\nonumber\\
m_T &> 1340~\GeV &&\text{for }T\to Zt,
\nonumber\\
m_T &> 1350~\GeV &&\text{for }T\to Wb.
\label{eq:representative-direct-limits}
\end{align}
These values characterize the sensitivity of conventional VLQ searches
in the stated benchmark scenarios. They do not imply that a vector-like
top partner below the quoted masses is excluded independently of its
decay pattern. The leading QCD pair-production cross section is governed
primarily by $m_T$ and the color representation of $T$, rather than by
its electroweak transition couplings. The resulting mass limits
nevertheless remain conditional on the assumed branching fractions,
fiducial acceptance, resonance width, and any additional non-QCD
production contributions.\\
The conventional decay modes $T\to Wb$, $T\to Zt$, and $T\to Ht$ do
not exhaust the possible phenomenology of vector-like quarks. In models
containing additional scalar or pseudoscalar states, decays such as
$T\to t\phi$, where $\phi$ denotes a new neutral boson, can generate
qualitatively different collider signatures
\cite{Cacciapaglia:2019zmj,Bhardwaj:2022wfz,CMS:2025zwi}. The
radiative modes $T\to t\gamma$ and $T\to tg$ considered in this work
constitute another class of nonstandard decays and motivate searches
complementary to both the conventional channels and dedicated
exotic-decay analyses.\\
Most experimental searches for vector-like top partners target the
renormalizable electroweak interactions responsible for the $Wb$, $Zt$,
and $Ht$ final states. More general phenomenological scenarios can
contain higher-dimensional transition interactions involving the heavy
quark, an SM quark, and a gauge-field strength
\cite{Tong:2023lms,Kim:2018mks,Alhazmi:2018whk}. For a charge-$2/3$
top partner, electromagnetic and chromomagnetic transitions generate
the decays $T\to t\gamma$ and $T\to tg$, producing energetic photons
and additional hard-jet activity. Strong-sector dynamics can provide a
setting for such interactions
\cite{Giudice:2007fh,Panico:2015jxa}, while explicit loop-induced
top-partner transitions have been studied in
Ref.~\cite{Kim:2018mks}. General VLQ frameworks relate the
electroweak representation and mixing structure of the heavy quark to
its conventional production and decay modes
\cite{delAguila:2000rc,Atre:2011ae}. The normalization of the
transition operators and their possible correlations with $TtZ$, $TtH$,
and $TbW$ interactions are therefore ultraviolet-model dependent. \\
Radiative decays and chromomagnetic interactions of vector-like top
partners have been investigated in dedicated phenomenological studies
\cite{Alhazmi:2018whk,Belyaev:2021zgq,Tong:2023lms}; related recent
work has also developed jet-substructure strategies for top-partner
searches \cite{Ghosh:2025vts}. In particular,
Ref.~\cite{Alhazmi:2018whk} studied a small-mixing regime in which the
conventional $Wb$, $Zt$, and $Ht$ decays are suppressed and radiative
decays can become dominant. Its electroweak-gauge-invariant hypercharge
operator produces correlated $Tt\gamma$ and $TtZ$ transitions and
assumes a definite chiral structure. The parametrization adopted in the
present work is not identical to that construction: it is formulated
directly after electroweak symmetry breaking, assumes no chiral
projection, treats the $Tt\gamma$ and $Ttg$ coefficients independently,
and does not include a correlated $TtZ$ interaction. The model is therefore formulated directly in terms of physical fields
after electroweak symmetry breaking. It respects colour and electromagnetic
gauge invariance but is not intended to constitute a complete
$SU(2)_L\times U(1)_Y$-invariant effective field theory. \\
A CMS resonance search for pair-produced spin-$1/2$ particles decaying
to $tg$ excludes masses below approximately $1.05~\TeV$ when
$\mathcal B(T\to tg)=1$ \cite{CMS:2024tgj}. This result is directly
relevant to the gluon-dominated limit of the radiative benchmark and is
shown for comparison with the final interpretation. It cannot, however,
be transferred without qualification to parameter points with a
nonzero $t\gamma$ branching fraction or with coupling-dependent
contributions to $T\bar T$ production.\\
In this work, we constrain the electromagnetic and chromomagnetic
transition interactions of a vector-like top partner using precision
measurements of top-quark production in association with photons.
Rather than performing a dedicated resonance search, we reinterpret the
unfolded photon-transverse-momentum spectrum in dileptonic
$t\bar t\gamma$ production measured by CMS \cite{CMS:2022lmh} and the
fiducial $t\bar t\gamma\gamma$ cross section measured by ATLAS
\cite{ATLAS:2025aps}. This strategy probes radiative decay
configurations that are not targeted directly by searches in the
conventional $Wb$, $Zt$, and $Ht$ channels and complements general
frameworks for interpreting VLQ searches
\cite{Buchkremer:2013bha,Matsedonskyi:2014mna,
Cacciapaglia:2011fx,Aguilar-Saavedra:2013qpa}.\\
The analysis adopts a radiative benchmark motivated by the
negligible-mixing limit. The conventional mixing-induced decays
$T\to Wb$, $T\to Zt$, and $T\to Ht$ are not included, and the optional
charged-current transition in the general model implementation is
switched off. The total width in the benchmark therefore contains only
the $T\to t\gamma$ and $T\to tg$ partial widths, and their physical
branching fractions satisfy
$\mathcal B(T\to t\gamma)+\mathcal B(T\to tg)=1$. This assumption
defines the two-transition model constrained in the present work; the
results are not intended to apply without modification to models
containing additional electroweak decay modes.\\
A controlled reinterpretation requires more than a comparison of
inclusive signal rates. The CMS likelihood and its validation use only
the published absolute $p_T^\gamma$ spectrum. Its signal prediction
contains the mixed
$T\bar T\to(t\gamma)(\bar t g)$ topology, including the
charge-conjugate assignment, together with
$T\bar T\to(t\gamma)(\bar t\gamma)$ events for which exactly one photon
satisfies the CMS fiducial definition. The latter contribution is
evaluated with its own topology-dependent acceptance rather than being
inferred from the mixed sample. The ATLAS result is treated as a
counting measurement with a signal-specific, mass-dependent correction
for the reconstruction response. Higher-order QCD normalization is
applied only to the pure-QCD component of $T\bar T$ production, while
renormalization- and factorization-scale, parton-distribution-function,
finite-simulation, migration, and response uncertainties are propagated
consistently through the statistical analysis.\\
The validity of the effective description is assessed separately at
every point in the coupling plane by removing EFT-dependent event
contributions whose hard scale exceeds the corresponding effective
scale. The additional requirement $\Gamma_T/m_T<0.1$ ensures that the
heavy state remains a well-defined resonance for which on-shell
production and subsequent decay can be factorized reliably, with
finite-width, nonresonant, and nonfactorizable effects remaining
controlled. Constraints from the CMS and ATLAS measurements are
reported separately because the information required to account
reliably for correlations between the two data sets is not publicly
available.\\
The remainder of this paper is organized as follows.
Section~\ref{sec:framework} introduces the effective transition interactions,
specifies their Lorentz structure, and presents the decay widths, branching
fractions, and decomposition of the production contributions. 
Section~\ref{sec:simulation} describes
the event generation and theory normalization.
Sections~\ref{sec:cms} and \ref{sec:atlas} present the CMS and ATLAS
reinterpretations, respectively. Section~\ref{sec:validity} specifies
the EFT and narrow-width validity requirements, and
Sec.~\ref{sec:statistics} defines the statistical treatment. The
results are presented in Sec.~\ref{sec:results}, and our conclusions
are summarized in Sec.~\ref{sec:conclusions}.

\section{Theoretical and phenomenological framework}
\label{sec:framework}

This section presents the theoretical framework used to describe the
interactions, production, and decay of the vector-like top partner. The
effective transition interactions are formulated directly in terms of
the physical photon and gluon fields after electroweak symmetry breaking.
The resulting description is a restricted broken-phase phenomenological
parametrization and is not assumed to constitute a complete
$SU(2)_L\times U(1)_Y$-gauge-invariant effective theory at energies
above the electroweak scale. Its relation to such a theory depends on
the electroweak representation and ultraviolet completion of the heavy
state. Within this explicitly defined framework, we derive the relevant
partial widths and physical branching fractions and construct the
coupling-dependent production and decay contributions entering the
fiducial signal predictions. The theoretical assumptions and conditions
required for the validity of the interpretation are stated explicitly.

\subsection{Effective dipole interactions}
\label{sec:eft}

We consider a vector-like Dirac fermion $T$ with electric charge $+2/3$
in the fundamental representation of $SU(3)_c$. For the ultraviolet
motivation of the neutral transition sector, it may be associated with
an $SU(2)_L$ singlet of hypercharge $Y=2/3$. Dipole transitions of
vector-like quarks have been studied, for example, in
Refs.~\cite{Tong:2023lms,Alhazmi:2018whk}. In the present work, we
consider the photon and gluon tensor vertices but no correlated
off-diagonal $TtZ$ vertex. The interaction is therefore formulated
directly in the mass-eigenstate electroweak-broken phase as
\begin{equation}
\mathcal L_{\rm int}
=
\sum_{i=1}^{3}\frac{f_i}{\Lambda}
\left[
g_\gamma\,\bar T\sigma^{\mu\nu}u_i F_{\mu\nu}
+
g_g\,\bar T\sigma^{\mu\nu}T^A u_i G^A_{\mu\nu}
\right]
+\mathrm{h.c.},
\label{eq:lagrangian}
\end{equation}
where
$\sigma^{\mu\nu}=i[\gamma^\mu,\gamma^\nu]/2$,
$u_i=(u,c,t)$, and $T^A$ are the fundamental $SU(3)_c$ generators.
The coefficients $f_i$ specify the flavor weights, while $g_\gamma$
and $g_g$ are dimensionless parameters controlling the electromagnetic
and chromomagnetic transitions, respectively. Thus,
$g_\gamma^i=f_i g_\gamma$ and $g_g^i=f_i g_g$ in the notation of
Ref.~\cite{Tong:2023lms}. Electromagnetic and strong gauge couplings are
not absorbed into the corresponding three-point transition
coefficients. The non-Abelian part of $G_{\mu\nu}^A$ nevertheless
generates the associated $Tqgg$ contact interaction proportional to
$g_s$, as required by QCD gauge invariance.\\
No chirality is assumed for the transition interactions in
Eq.~\eqref{eq:lagrangian}. In particular, the implemented Lorentz
structures contain neither explicit $P_L$ or $P_R$ projectors nor
independently adjustable left- and right-handed coefficients. The
interaction therefore corresponds to the unprojected tensor bilinears
written in Eq.~\eqref{eq:lagrangian}. The limits derived below apply to
this nonchiral phenomenological parametrization and should not be
interpreted as separate constraints on left- or right-handed dipole
operators. A translation to a model with a specified chiral structure
would require matching its operator normalization and, where relevant,
reevaluating chirality-dependent decay and acceptance effects.\\
For the third generation, the dimensionful transition coefficients used
in the interpretation are
\begin{equation}
c_{t\gamma}=\frac{f_3g_\gamma}{\Lambda},
\qquad
c_{tg}=\frac{f_3g_g}{\Lambda},
\label{eq:effectivecouplings}
\end{equation}
and are quoted throughout in $\GeV^{-1}$. The two coefficients are
varied independently in the statistical analysis.\\
We work in a radiative benchmark motivated by the limit of negligible
mixing between $T$ and the SM top quark. In this limit, the conventional
mixing-induced decays $T\to Wb$, $T\to Zt$, and $T\to Ht$ are
suppressed \cite{Alhazmi:2018whk}. The optional charged-current
transition present in the more general model implementation is switched
off and does not enter the Lagrangian, decay width, or signal prediction
used in this analysis. The resulting benchmark contains only the
$Tt\gamma$ and $Ttg$ transition interactions displayed in
Eq.~\eqref{eq:lagrangian}.\\
Equation~\eqref{eq:lagrangian} is invariant under
$SU(3)_c\times U(1)_{\rm em}$, but it is not, by itself, a complete
$SU(2)_L\times U(1)_Y$-invariant EFT. A charge-$2/3$ vector-like state
with the same gauge quantum numbers as $t_R$ can mix with the top quark
and thereby acquire the conventional $Wb$, $Zt$, and $Ht$
interactions. Transition dipoles may also be generated by compositeness,
heavy mediators, or loop-induced electromagnetic and chromomagnetic
moments. None of these specific ultraviolet mechanisms is selected by
Eq.~\eqref{eq:lagrangian}.\\
An electroweak-gauge-invariant completion above the symmetry-breaking
scale would be written in terms of the hypercharge and weak-isospin
field strengths and, where required by the field representations,
insertions of the Higgs doublet. After electroweak symmetry breaking,
such operators generally produce correlated photon and $Z$
transitions. Their relative coefficients depend on the electroweak
representation of $T$ and on the ultraviolet operator basis.
Consequently, retaining a broken-phase $Tt\gamma$ interaction while
omitting an off-diagonal $TtZ$ vertex is a deliberate phenomenological
restriction, not a generic consequence of electroweak gauge invariance.
In particular, negligible mixing suppresses the conventional
mixing-induced electroweak decays but does not, by itself, eliminate
a dipole-induced $TtZ$ interaction. Its omission is an independent
assumption defining the two-channel radiative benchmark studied here.\\
This restriction must also be distinguished from particular ultraviolet
realizations. In the partial-compositeness construction studied in
Ref.~\cite{Belyaev:2021zgq}, for example, the off-diagonal
chromomagnetic interaction arises after mixed elementary and composite
gauge eigenstates are rotated to the mass basis. Reducing the mixing
responsible for the conventional $Wb$, $Zt$, and $Ht$ interactions can
then also reduce the transition dipole. Other ultraviolet constructions
can produce different relations. In the gauge-invariant construction of
Ref.~\cite{Alhazmi:2018whk}, a hypercharge dipole generates correlated
$Tt\gamma$ and $TtZ$ interactions even when the conventional
mixing-induced decays are suppressed. We do not combine these
ultraviolet mechanisms into a single model. The limits are quoted for
the two vertices present in Eq.~\eqref{eq:lagrangian}; applying them to
a specific ultraviolet completion requires a dedicated matching
calculation and the inclusion of any additional correlated decay modes.\\
Reference~\cite{Tong:2023lms} considers nonzero first- and
second-generation coefficients, allowing single-resonance production
through $qg\to T$, followed by $T\to qg$ or $T\to q\gamma$. That
production regime is not adopted here. We impose $f_1=f_2=0$ as a
flavor benchmark, thereby removing the corresponding valence-quark
single-production channels and restricting the phenomenology to
third-generation transitions. This assumption is not inferred from a
model-independent flavor constraint on the coefficients in
Eq.~\eqref{eq:lagrangian}. Vector-like quarks coupled to light quarks
constitute a distinct phenomenological scenario
\cite{Okada:2012gy,Cacciapaglia:2017gzh,CMS:2017pvr}; constraints on
that scenario cannot be transferred directly to the present benchmark
without matching its flavor structure, production mechanisms, and
operator normalization. General flavor considerations
\cite{Isidori:2010kg} motivate stating the flavor assumption explicitly
but do not uniquely determine the vector-like-quark dipole coefficients.
The present analysis considers $T\bar T$ production followed by
third-generation radiative decays.

\subsection{Widths and physical branching fractions}
\label{sec:widths}

For the normalization in Eq.~\eqref{eq:lagrangian}, the tree-level
two-body partial widths for the electromagnetic and chromomagnetic
transitions $T \to t+g$ and $T\to t+\gamma$ are
\begin{equation}
\Gamma_{t\gamma}
=
\frac{m_T^3}{2\pi}
\left(1-\frac{m_t^2}{m_T^2}\right)^3
|c_{t\gamma}|^2,
\qquad
\Gamma_{tg}
=
\frac{2m_T^3}{3\pi}
\left(1-\frac{m_t^2}{m_T^2}\right)^3
|c_{tg}|^2.
\label{eq:radiativewidths}
\end{equation}
For equal values of the two dimensionful coefficients,
$\Gamma_{tg}/\Gamma_{t\gamma}=4/3$, reflecting the color factor of a
fundamental quark. The cubic dependence on $m_T$ follows from the
momentum dependence of the dimension-five transition interaction, while
the factor $(1-m_t^2/m_T^2)^3$ accounts for the finite top-quark mass.\\
The signal samples use factorized on-shell production and decay. The
physical interpretation is therefore restricted to
$\Gamma_T/m_T<0.1$, for which $T$ remains a well-defined resonance and
corrections to the narrow-width treatment are expected to remain
controlled. This upper-width requirement excludes broad-resonance
regions in which production and decay cannot be factorized reliably. It
does not imply that finite-width, nonresonant, and nonfactorizable
effects vanish identically, but restricts the interpretation to the
region in which they are expected to remain parametrically controlled.
The prompt-decay assumption is well justified throughout the reported
parameter region. For illustration, at $m_T=500~\GeV$, taking
$c_{t\gamma}=10^{-5}~\GeV^{-1}$ and $c_{tg}=0$ gives
$c\tau_T\simeq1.45\times10^{-13}~\mathrm{m}$, far below detector
resolution scales. Larger couplings or an additional nonzero
chromomagnetic coupling increase the total width and further shorten
the decay length.

\subsection{Production decomposition}
\label{sec:production}

\begin{figure}[ht]
\centering
\includegraphics[width=0.650\textwidth]{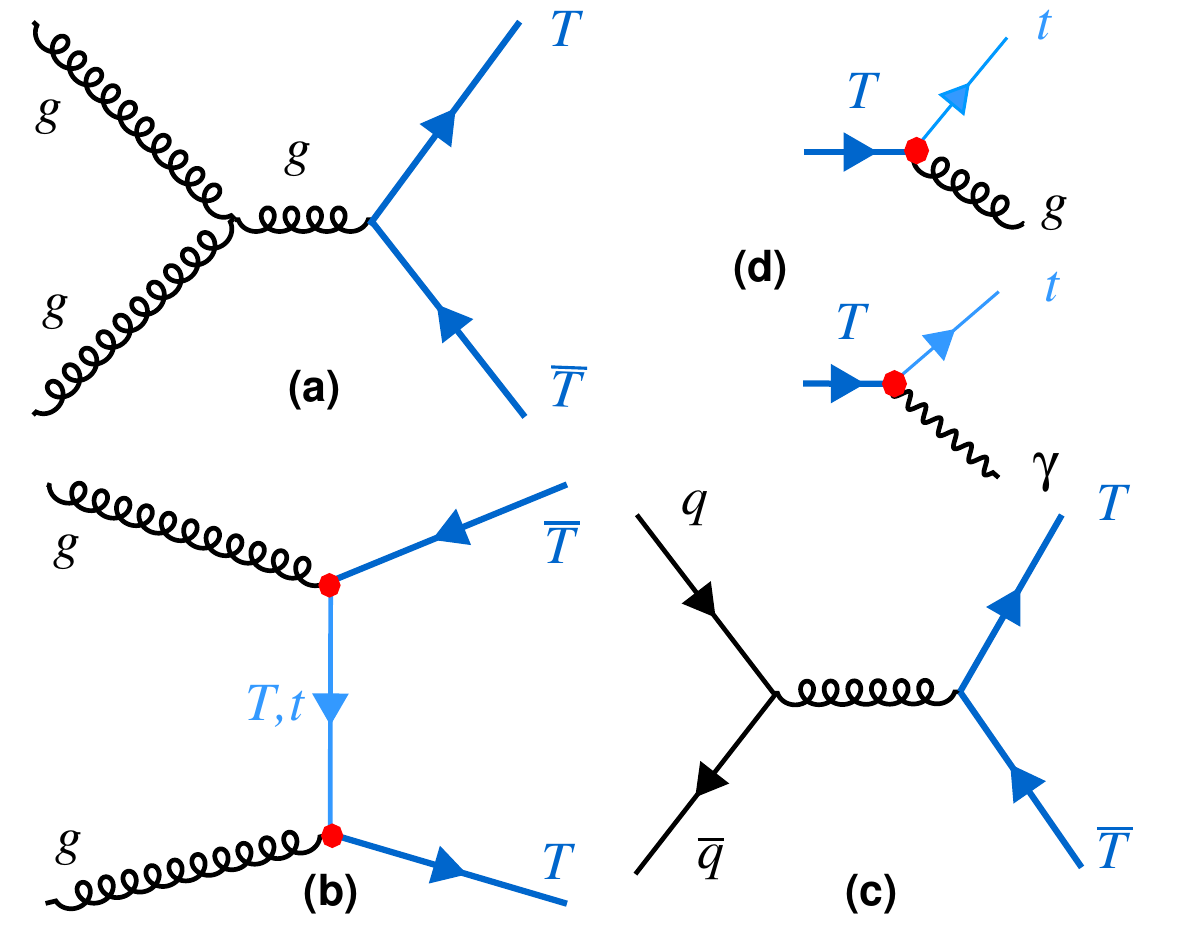}
\caption{Representative production and decay diagrams for the
vector-like top partner. Panel (a) shows the $s$-channel gluon
contribution to $gg\to T\bar T$, and panel (c) shows
quark--antiquark annihilation; both belong to pure-QCD production.
Panel (b) represents a $t$-channel topology in gluon fusion. With
ordinary $gT\bar T$ vertices it contains heavy-$T$ exchange, whereas
two chromomagnetic $Ttg$ vertices generate the top-exchange transition
amplitude. Panel (d) illustrates the decays $T\to tg$ and
$T\to t\gamma$. Crossed $u$-channel diagrams, charge-conjugate
processes, the remaining QCD diagrams, and all corresponding
interference terms are included in the matrix-element calculation
although they are not all displayed. Red vertices denote insertions of
the $Ttg$ transition interaction.}
\label{fig:feynmandiagrams}
\end{figure}

At hadron colliders, vector-like top partners are produced in pairs
predominantly through strong interactions,
$pp\to T\bar T$. In the presence of the $Ttg$ transition vertex in
Eq.~\eqref{eq:lagrangian}, the pair-production cross section can be
decomposed as
\begin{equation}
\sigma_{T\bar T}
=
\sigma_{\rm QCD}
+c_{tg}^{\,2}\sigma_{\rm int}
+c_{tg}^{\,4}\sigma_{\rm dip}.
\label{eq:decomposition}
\end{equation}
The dimensions of $\sigma_{\rm int}$ and $\sigma_{\rm dip}$ compensate
the explicit powers of $c_{tg}$.
The coupling powers in Eq.~\eqref{eq:decomposition} follow directly
from the amplitude structure. At tree level, the complete
$gg\to T\bar T$ matrix element contains the QCD $s$-channel diagram,
the QCD $t$- and $u$-channel diagrams with heavy-$T$ exchange, and
transition-induced $t$- and $u$-channel diagrams with top-quark
exchange. The $q\bar q\to T\bar T$ contribution is included separately.
The transition-induced gluon-fusion amplitude contains two $Ttg$
insertions and is therefore proportional to $c_{tg}^2$. Its
interference with the QCD amplitude is proportional to $c_{tg}^2$, and
its squared contribution is proportional to $c_{tg}^4$. The
interference coefficient can have either sign after phase-space
integration and can vary across the kinematic phase space; it must not
be interpreted as an independently positive event yield.\\
The dominant production mechanisms and decay transitions are illustrated
in Fig.~\ref{fig:feynmandiagrams}. The prediction is obtained from the
complete set of tree-level matrix elements, including the crossed
channels, charge-conjugate processes, and all remaining QCD topologies.
The calculation therefore does not rely on any single representative
diagram.

The momentum carried by the field strength enhances the relative
importance of the transition-induced contribution at large partonic
energies. The dipole component can therefore become increasingly
important in the high-$M_{T\bar T}$ and high-transverse-momentum tails.
The same energy enhancement that improves experimental sensitivity also
drives the amplitude toward the boundary of the effective description.
For this reason, the $M_{T\bar T}$ validity requirement is applied
event by event and separately in each analysis bin, as described in
Sec.~\ref{sec:validity}; it cannot be represented by a single inclusive
efficiency.\\
The gluon or photon produced in a two-body decay of a heavy $T$ quark
typically carries a substantial fraction of the parent-particle energy.
The radiative decays therefore give rise to energetic photons or jets,
providing sensitivity in the high-transverse-momentum regions of the
measurements considered below.

\subsection{Decay topologies relevant to the photon measurements}
\label{sec:decaytopologies}

Pair production followed by the radiative decays produces three final-state configurations. The
double-radiative topology
\begin{equation}
T\bar T\to(t\gamma)(\bar t\gamma)
\label{eq:gammagammatopology}
\end{equation}
has branching-fraction weight $\mathcal B_\gamma^2$ and is the signal
topology directly probed by the ATLAS $t\bar t\gamma\gamma$
measurement. The same topology can also enter the CMS
exactly-one-fiducial-photon selection when only one of the two generated
photons satisfies the published fiducial requirements. This migration
is evaluated from an independent double-radiative sample and is not
represented by an inclusive rescaling of the mixed topology.\\
The mixed photon--gluon topology
\begin{equation}
T\bar T\to(t\gamma)(\bar t g)
\quad\text{and its charge-conjugate assignment}
\label{eq:gammagtopology}
\end{equation}
has branching-fraction weight
$2\mathcal B_\gamma\mathcal B_g$ and provides the principal
single-photon contribution to the CMS reinterpretation. The factor of
two accounts for the two assignments of the photon and gluon decay modes
to $T$ and $\bar T$. The gluon produced in the heavy-quark decay gives
rise to an additional energetic hadronic jet.\\
The heavy-$T$ contribution to the CMS single-photon fiducial cross section
in bin $i$ is therefore written as
\begin{equation}
s_i^{\rm CMS}
=
s_{i,\gamma g}^{(1\gamma_{\rm fid})}
+
s_{i,\gamma\gamma}^{(1\gamma_{\rm fid})},
\label{eq:cmssignaltopologies}
\end{equation}
where the superscript identifies events satisfying the
exactly-one-fiducial-photon requirement. The two contributions are evaluated
using independent event samples and normalized with the physical
branching-fraction weights
$2\mathcal B_\gamma\mathcal B_g$ and $\mathcal B_\gamma^2$,
respectively. The second term describes generated double-radiative events
for which only one photon satisfies the fiducial requirements. It therefore
accounts for migration into the single-photon category, rather than an
additional decay mode or a rescaling of the mixed-topology contribution.\\
The remaining radiative configuration,
$T\bar T\to(tg)(\bar t g)$, has branching-fraction weight
$\mathcal B_g^2$ but contains no prompt photon from a $T$ decay. It
therefore does not contribute to the signal definition used in either
photon measurement. This topology-dependent construction keeps the
production rate, physical branching fractions, and fiducial migration
distinct throughout the analysis.

\section{Simulation and prediction construction}
\label{sec:simulation}
Events are generated with \textsc{MadGraph5\_aMC@NLO} v3.5.12
\cite{Alwall:2011uj,Alwall:2014hca} using a UFO implementation prepared with
\textsc{FeynRules} \cite{Alloul:2013bka,Degrande:2011ua} and following the UFO
format conventions \cite{Darme:2023jdn}. The nominal parton distributions
belong to the NNPDF3.0 family \cite{Ball:2014uwa}. 
Showering and hadronization
use \textsc{Pythia}~8.3; anti-$k_T$ jets are clustered
with the algorithm and software framework of Refs.~\cite{Cacciari:2008gp,Cacciari:2011ma} and parameter choices matched
to each measurement. 
Separate samples are generated for the production components and decay topologies required to reconstruct Eq.~\eqref{eq:decomposition}.  
The reference signal samples use
$ m_T=500,750,\ldots,3000~\GeV,$ and  $\Lambda=5000~\GeV,$
with a mass spacing of $250~\GeV$.  The normalization
$\Lambda=5~\TeV$ used to define the vertices is distinct from the
point-dependent diagnostic $\Lambda_{\rm eff}$ used to ensure the EFT validity which will be described in Sec.~\ref{sec:validity}. 
$\Lambda_{\rm eff}$ is constructed from the physical dimensionful coefficients and is used only for the event-level validity test.
The coefficient templates are obtained either from dedicated samples or from a polynomial reweighting 
validated at independent coupling points. In either case, the closure test is required to reproduce total rates and bin contents 
at couplings not used to determine the polynomial. \\
Dedicated NLO calculations of vector-like quark pair production have
demonstrated that QCD radiative corrections significantly increase the
inclusive production cross section while only moderately affecting the
overall event kinematics~\cite{Fuks:2016ftf}. Motivated by these
results, the pure-QCD production component is normalized using a
constant inclusive QCD correction factor,
$K_{\rm QCD}=1.40$ \cite{Fuks:2016ftf},
which is representative of the NLO-to-LO correction over the
vector-like top mass range considered in this analysis.
Since no dedicated higher-order calculation is available for the
interference between the SM QCD amplitudes and the dipole
operators, or for the pure dipole-induced production contribution, no
additional $K$ factors are applied to these terms. The predicted signal
cross section is therefore written as
\begin{equation}
\sigma_{T\bar T}^{\rm pred}
=
K_{\rm QCD}\,
\sigma_{\rm QCD}^{\rm LO}
+
c_{tg}^{\,2}\,
\sigma_{\rm int}^{\rm LO}
+
c_{tg}^{\,4}\,
\sigma_{\rm dip}^{\rm LO},
\label{eq:kfactor}
\end{equation}
where the first term denotes the QCD pair-production contribution,
while the second and third terms correspond to the QCD--EFT
interference and the pure EFT contribution, respectively.
Theoretical uncertainties in the signal prediction are estimated using
the standard seven-point renormalization- and factorization-scale
variations, excluding the combinations in which one scale is doubled
and the other is halved, together with the PDF uncertainties provided
by the chosen parton distribution function set. These variations are
propagated through the full event selection and incorporated as
correlated systematic uncertainties in the statistical interpretation.

\subsection{Object definitions and fiducial implementation}
Stable-particle lifetimes, lepton dressing, prompt-photon ancestry, photon isolation, jet inputs, heavy-flavor labeling, 
and overlap removal are implemented separately for CMS and ATLAS. These details are physically relevant for the signal. 
A photon from $T$ decay is prompt, but nearby radiation from the boosted top system can alter isolation. 
The hard gluon in the mixed topology changes jet multiplicity and can overlap a photon or lepton. 
At large $m_T$, boosted top decay products can approach the fixed-radius object-separation boundaries. 
Acceptances are consequently extracted from the complete event record
rather than assumed to be constant or inherited from SM
$t\bar t\gamma$ or $t\bar t\gamma\gamma$ production.
Two response levels are maintained: particle-level fiducial acceptance and reconstruction-level efficiency. CMS is compared directly to an unfolded particle-level spectrum after validation. ATLAS requires the additional signal-specific transfer described in \Cref{sec:atlas}. This separation avoids using detector effects twice.

\section{CMS single-photon reinterpretation}
\label{sec:cms}
We use only the absolute unfolded $p_T^\gamma$ spectrum in the CMS dilepton fiducial region at $\sqrt{s}=13~\TeV$ and $138~\fb^{-1}$ \cite{CMS:2022lmh}. 
The phrase ``CMS single-photon channel'' denotes the particle-level fiducial
category used in the unfolded measurement; it does not identify a unique
parton-level decay topology.  
The particle-level fiducial phase space is defined by the following
requirements \cite{CMS:2022lmh}:
\begin{itemize}
    \item Two oppositely charged leptons ($e$ or $\mu$) with
          $p_{T,\ell_1} > 25\;\GeV$, $p_{T,\ell_2} > 15\;\GeV$,
          $|\eta_\ell| < 2.4$, and dilepton invariant mass
          $m_{\ell\ell} > 20\;\GeV$;
    \item At least one jet with $p_{T,j} > 30\;\GeV$, $|\eta_j| < 2.4$,
          matched to a $b$-hadron within $\Delta R < 0.4$;
    \item One isolated photon with $p_T^\gamma > 20\;\GeV$ and $|\eta_\gamma| < 1.44$, required to be separated from any other stable particle with $p_T > 5\;\GeV$ (except neutrinos) by $\Delta R > 0.1$;
       \item Angular separation $\Delta R(\ell,\gamma) > 0.4$,
          $\Delta R(j,\gamma) > 0.1$, $\Delta R(\ell,j) > 0.4$.
\end{itemize}
The photon transverse momentum distribution provides a direct probe of
the energy scale associated with the dipole interaction.
The mixed decay configuration $T\bar T\to(t\gamma)(\bar t g)+(t g)(\bar t\gamma)$ naturally contains one
photon from a heavy-particle decay.  The double-radiative configuration
$T\bar T\to(t\gamma)(\bar t\gamma)$ instead contains two prompt photons before
fiducial requirements, but it can still populate an exactly-one-photon category
when one photon fails the particle-level photon definition.  The two topologies
are generated and selected independently because their branching weights,
object multiplicities, overlap-removal behavior, kinematic distributions, and
selection acceptances are different.  As mentioned previously, their accepted bin yields are added only
after the full selection:
\begin{equation}
s_i^{\rm CMS}=s_{i,\gamma g}^{(1\gamma_{\rm fid})}
+s_{i,\gamma\gamma}^{(1\gamma_{\rm fid})}. \nonumber
\end{equation}
The second term describes migration of a generated double-radiative
topology into the exactly-one-photon category; it does not represent a new decay
mode, a disappearing photon, or an arbitrary rescaling of the mixed sample.
Events in which both decay photons satisfy the CMS particle-level photon requirements are
excluded from the CMS prediction.  Events in which exactly one photon satisfies
them are retained once, and the transverse momentum of that unique fiducial
photon determines bin $i$.  An event with neither photon in the fiducial region
does not contribute.  This exclusive classification prevents both omission of
a physical signal component and double counting of a diphoton event.

\subsection{Migration of the double-radiative topology}
\label{sec:cmsmigration}

\begin{figure}[!ht]
\centering
\includegraphics[width=0.49\textwidth]
{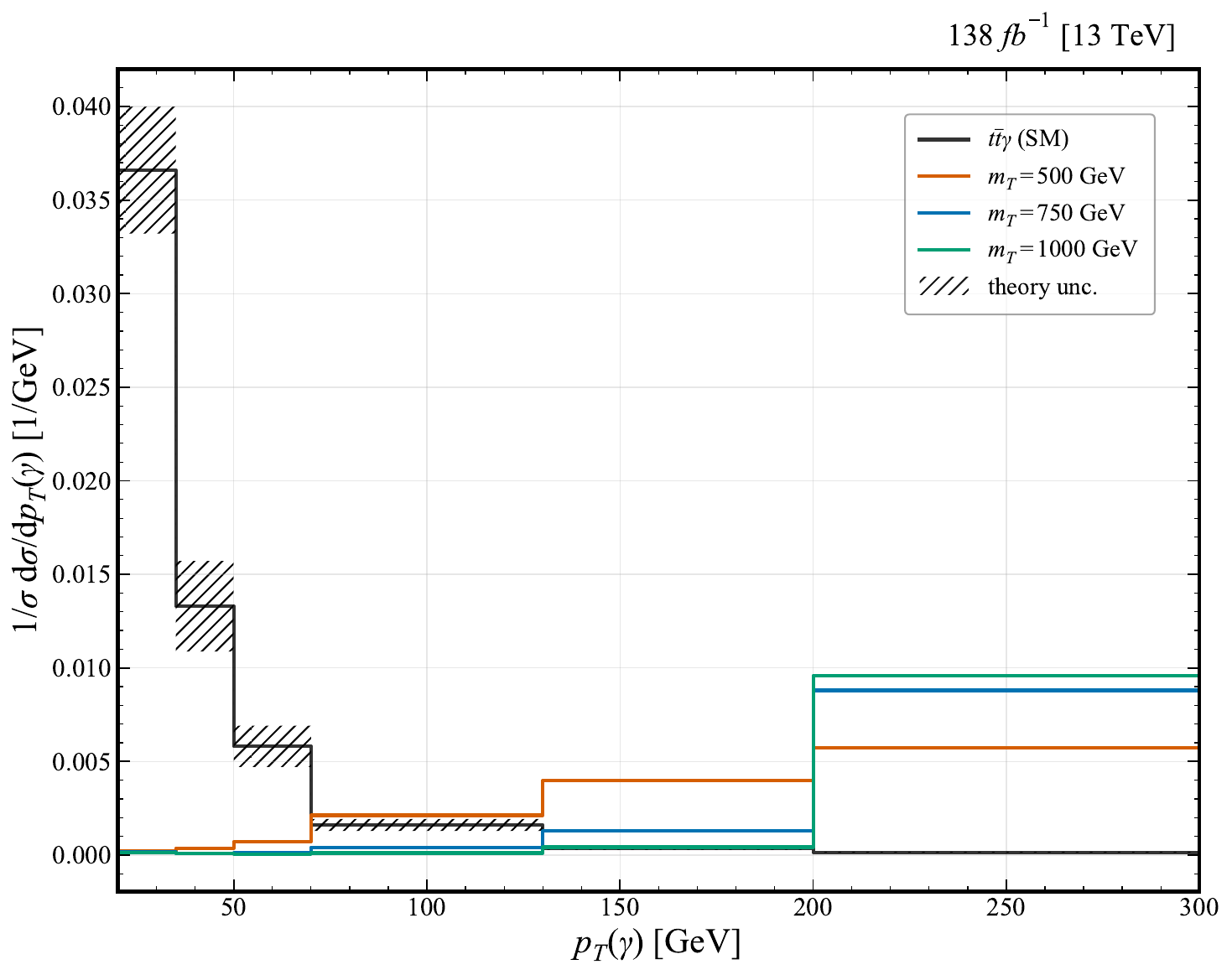}
\hfill
\includegraphics[width=0.49\textwidth]
{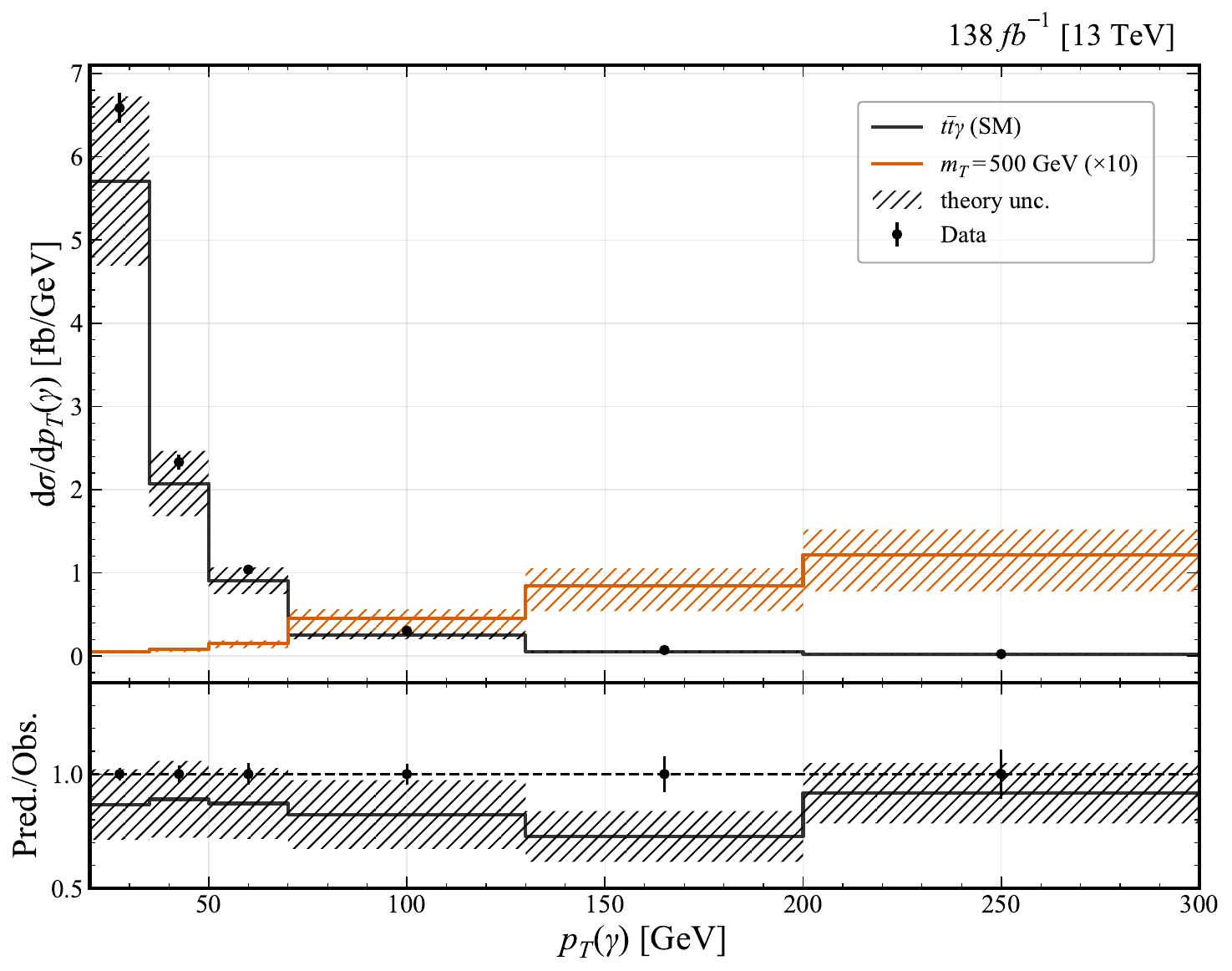}
\caption{Particle-level photon transverse-momentum distributions in the
dileptonic $t\bar t\gamma$ fiducial region at $\sqrt{s}=13~\TeV$.
Left: normalized distributions,
$(1/\sigma_{\rm fid})\,d\sigma_{\rm fid}/dp_T^\gamma$, for the SM
prediction and heavy-$T$ signals with $m_T=500$, $750$, and
$1000~\GeV$, illustrating their different spectral shapes.
Right: absolute differential fiducial cross sections for the SM prediction,
the CMS data corresponding to $138~\fb^{-1}$ \cite{CMS:2022lmh},
and the heavy-$T$ signal with $m_T=500~\GeV$. The signal is shown
separately from the SM contribution and multiplied by ten solely for
visibility. All signal benchmarks use
$c_{t\gamma}=c_{tg}=1.0\times10^{-4}~\GeV^{-1}$ and include both
the mixed $(t\gamma)(tg)$ topology and double-radiative
$(t\gamma)(t\gamma)$ events satisfying the exactly-one-fiducial-photon
requirement, with their respective physical branching-fraction weights.
Hatched bands indicate theoretical uncertainties, while the data points
show the measured values and their uncertainties. The lower panel of the
absolute distribution shows the SM prediction divided by the measured
cross sections. The likelihood uses the absolute spectrum and the
unscaled signal prediction.}
\label{fig:differential_spectra}
\end{figure}

Besides the mixed decay topology
$T\bar T\rightarrow (t\gamma)(\bar tg)$, the double-radiative channel
$T\bar T\rightarrow (t\gamma)(\bar t\gamma)$ also contributes to the
CMS single-photon measurement whenever exactly one of the two photons
satisfies the fiducial selection. At particle level, the second photon may
fail the selection because its transverse momentum falls below threshold,
its pseudorapidity lies outside the accepted region, it fails the isolation
requirement, or it is removed by the prescribed overlap criteria with a
dressed lepton or jet. The probability of this migration depends on the
heavy-quark mass, the event kinematics, additional QCD radiation, and the
top-decay topology, and is therefore determined by applying the complete
CMS fiducial selection to fully simulated double-radiative events. It is
not approximated by a constant photon inefficiency. \\
This particle-level migration is conceptually distinct from detector-response
migration. The published CMS measurement is unfolded to an exactly-one-photon
particle-level fiducial phase space, and the signal prediction is therefore
constructed at the same particle level. Detector effects are already accounted
for through the experimental response matrix and are not modeled separately in
the theoretical prediction.\\
For each of the two photons, indexed by $k=1,2$, let
$I_\gamma(\gamma_k)$ equal unity if that photon satisfies the CMS
fiducial requirements and zero otherwise. The indicator for exactly
one of these photons satisfying the requirements is
\begin{equation}
I_{1\gamma}
=
I_\gamma(\gamma_1)\left[1-I_\gamma(\gamma_2)\right]
+
I_\gamma(\gamma_2)\left[1-I_\gamma(\gamma_1)\right].
\label{eq:onephotonindicator}
\end{equation}
This expression equals unity when exactly one photon passes and vanishes
when both or neither pass. The accepted photon is therefore uniquely
identified, independently of their ordering in the event record.
The signal contributions entering CMS photon-$p_T$ bin $i$ are
\begin{align}
s_{i,\gamma\gamma}^{(1\gamma_{\rm fid})}
&=
\mathcal{L}\,
\sigma_{T\bar T}\,
\mathcal{B}_\gamma^2\,
A_{i,\gamma\gamma}^{(1\gamma_{\rm fid})},
\label{eq:doublemigrationyield}
\\
s_{i,\gamma g}^{(1\gamma_{\rm fid})}
&=
\mathcal{L}\,
\sigma_{T\bar T}\,
2\mathcal{B}_\gamma\mathcal{B}_g\,
A_{i,\gamma g}^{(1\gamma_{\rm fid})},
\label{eq:mixedyield}
\end{align}
where
$A_{i,X}^{(1\gamma_{\rm fid})}
=
\frac{
N_{i,X}^{(1\gamma_{\rm fid})}
}{
N_{X}^{\rm generated}
}, X\in\{\gamma g,\gamma\gamma\}, $
is the fiducial acceptance for topology $X$, defined as the fraction of
generated events that satisfy the complete CMS particle-level fiducial
selection with exactly one accepted photon and are assigned to
photon-$p_T$ bin $i$. Consequently, the acceptance includes the
branching fractions of the dileptonic top decays, all fiducial object
requirements, the exactly-one-photon condition, and the assignment to
bin $i$. Whenever the production kinematics differ, separate
acceptances are evaluated for the pure-QCD, interference, and
dipole-squared production components. The total BSM prediction in each
photon-$p_T$ bin is then obtained by summing the two contributions,
$s_i^{\rm BSM} = s_{i,\gamma g}^{(1\gamma_{\rm fid})} + s_{i,\gamma\gamma}^{(1\gamma_{\rm fid})}.$
The fractional contribution of the migrated double-radiative topology is
\begin{equation}
R_{\gamma\gamma,i}^{\rm CMS}
=
\frac{\mathcal{B}_\gamma^2
A_{i,\gamma\gamma}^{(1\gamma_{\rm fid})}}
{2\mathcal{B}_\gamma\mathcal{B}_g
A_{i,\gamma g}^{(1\gamma_{\rm fid})}
+
\mathcal{B}_\gamma^2
A_{i,\gamma\gamma}^{(1\gamma_{\rm fid})}},
\label{eq:migrationfraction}
\end{equation}
which varies over the EFT parameter space through the physical branching
fractions and, to a lesser extent, through the acceptance.\\
The migration study was performed for every benchmark mass considered
in this analysis. For
$c_{t\gamma}=c_{tg}=1.0\times10^{-4}\,\GeV^{-1}$, the
double-radiative topology contributes approximately $11$--$14\%$
of the total selected CMS single-photon signal, corresponding to an
increase of about $12$--$16\%$ relative to the mixed
$(t\gamma)(tg)$ contribution alone. Specifically, this fraction is
$13.7\%$ at $m_T=500\,\GeV$, reaches a minimum of $10.9\%$
around $m_T=2000\,\GeV$, and increases slightly to $12.3\%$
at $m_T=3000\,\GeV$. The weak mass dependence refers to this
benchmark; throughout the coupling scan, the relative contribution
is evaluated using the physical branching fractions and
topology-dependent acceptances at each tested point.
This weak mass dependence demonstrates that the migration is an
intrinsic feature of the event topology rather than an accidental effect at a
particular benchmark point. Consequently, the double-radiative contribution is
included consistently in the signal prediction for every heavy-quark mass and
every photon-$p_T$ bin. An ablation study, in which this contribution is
removed while keeping the statistical procedure and systematic uncertainties
unchanged, is performed to quantify its impact on the expected and observed
exclusion limits.\\
Fig.~\ref{fig:differential_spectra} illustrates the photon
transverse-momentum spectra entering the CMS reinterpretation. The normalized
distributions compare the SM prediction with heavy-$T$ signals at
$m_T=500$, $750$, and $1000~\GeV$, emphasizing their different kinematic
shapes. The absolute differential distribution compares the SM prediction
with the unfolded CMS data and shows the heavy-$T$ contribution for
$m_T=500~\GeV$ separately. All signal benchmarks use
$c_{t\gamma}=c_{tg}=1.0\times10^{-4}~\GeV^{-1}$.
The SM prediction is obtained using the simulation and particle-level
fiducial implementation described previously, with the theoretical
uncertainties shown bin by bin. Its comparison with the measured absolute
spectrum provides a validation of the fiducial prediction used in the
reinterpretation.\\
The heavy-$T$ prediction includes the mixed
$T\bar T\to(t\gamma)(\bar t g)$ topology, together with its
charge-conjugate assignment, and double-radiative
$T\bar T\to(t\gamma)(\bar t\gamma)$ events satisfying the
exactly-one-fiducial-photon requirement. These contributions are evaluated
independently and normalized using their corresponding physical
branching-fraction weights. In the absolute distribution, the
$m_T=500~\GeV$ signal is multiplied by ten solely for visibility;
this display factor is not used in the statistical analysis.\\
The normalized spectra demonstrate that photons from heavy-$T$ decays
populate a substantially harder distribution than the SM contribution,
with the spectrum shifting toward higher $p_T^\gamma$ as $m_T$
increases. This behavior reflects the increasing photon energy available
in the two-body decay $T\to t\gamma$. The high-$p_T^\gamma$ bins
therefore provide important discrimination between the signal and SM
predictions, motivating a differential analysis rather than an
inclusive-rate interpretation. The likelihood uses only the absolute
$d\sigma/dp_T^\gamma$ measurement, its associated covariance matrix,
and the unscaled signal prediction in every bin to determine the
constraints in the $(c_{tg},c_{t\gamma})$ plane. The normalized
distributions are shown only to illustrate the spectral differences and
are not used in the limit extraction.

The complete CMS reinterpretation is validated through a series of closure tests.
The particle-level implementation reproduces the published SM
fiducial normalization and unfolded $p_T^\gamma$ spectrum using the same
particle-level definition, including the published overflow convention for the
last photon-$p_T$ bin. Agreement is quantified using the experimental covariance
matrix,
\begin{equation}
\chi^2_{\rm closure}
=
(\mathbf{m}-\mathbf{r})^T
(V_{\rm CMS}+V_{\rm MC})^{-1}
(\mathbf{m}-\mathbf{r}),
\end{equation}
where $\mathbf{m}$ and $\mathbf{r}$ denote the reproduced and published CMS
predictions, respectively, while $V_{\rm MC}$ represents the
simulation-statistical covariance matrix. The resulting agreement,
$\chi^2/\mathrm{NDF}=1.03$, demonstrates excellent reproduction of the CMS
particle-level prediction.

\section{ATLAS diphoton reinterpretation}
\label{sec:atlas}
The ATLAS measurement of $t\bar t\gamma\gamma$ production at
$\sqrt{s}=13~\TeV$, based on $140~\fb^{-1}$ of Run~2 data, is used as a
single fiducial counting observable \cite{ATLAS:2025aps}. The measured
fiducial cross section and the corresponding SM prediction
entering the reinterpretation are
\begin{equation}
 \sigma_{\rm obs}=2.51~\fb,
 \qquad \Delta\sigma_{\rm obs}=0.60~\fb,
 \qquad
 \sigma_{\rm SM}=1.53~\fb,
 \qquad \Delta\sigma_{\rm SM}=0.46~\fb.
 \label{eq:atlasinputs}
\end{equation}
No unpublished differential distribution, bin-to-bin covariance, or
event-category breakdown is used or inferred: the reinterpretation is
constructed from exactly the information ATLAS makes public, so that it can
be reproduced independently.\\
At particle level, photons are required to be prompt and non-hadronic, to
satisfy $p_T^\gamma>20~\GeV$ and $|\eta^\gamma|<2.37$, and to obey the
charged-particle isolation requirement $\sum_{\Delta R\leq0.2}p_T^{\rm
ch}<0.05\,p_T^\gamma$. Leptons are dressed with non-hadronic photons within
$\Delta R=0.1$ and must satisfy $p_T^\ell>25~\GeV$ and $|\eta^\ell|<2.5$.
Jets are reconstructed with the anti-$k_t$ algorithm ($R=0.4$) and required
to have $p_T^j>25~\GeV$ and $|\eta^j|<2.5$. The fiducial selection requires
exactly two photons, exactly one electron or muon, at least four jets, and
at least one ghost-matched $b$ jet; events with an additional lepton of
$p_T>7~\GeV$ are vetoed. Photon--photon and photon--lepton pairs are required
to satisfy $\Delta R>0.4$, and the prescribed jet--photon and jet--lepton
overlap removal is applied throughout.

\subsection{Mass-dependent detector-response transfer factor}
\label{sec:correction}

The decay products of a heavy $T\bar T$ pair are substantially harder than
those of the SM $t\bar t\gamma\gamma$ process that ATLAS measured, and the
resulting topology becomes increasingly boosted as $m_T$ grows. Because
photon identification and isolation, resolved-jet multiplicity, $b$-tagging,
and object-overlap removal are all sensitive to how collimated and how busy
the final state is, the reconstruction-to-fiducial response measured by
ATLAS for the SM process cannot be assumed to hold for the signal. Applying
the SM efficiency directly to the signal would therefore introduce a
mass-dependent bias into the inferred fiducial signal yield.\\
To quantify and account for this response difference, the SM validation sample and every signal
benchmark are passed through an identical simulation and analysis
chain, at both particle and reconstruction level. We therefore define the
mass-dependent detector-response transfer factor,
\begin{equation}
C(m_T)=
\frac{\left(N_{\rm reco}/N_{\rm fid}\right)_{T\bar T}}
     {\left(N_{\rm reco}/N_{\rm fid}\right)_{t\bar t\gamma\gamma}^{\rm Delphes}},
\label{eq:correction}
\end{equation}
where $N_{\rm fid}$ is the number of events satisfying the particle-level
fiducial selection above and $N_{\rm reco}$ the number surviving the
corresponding selection applied to reconstructed objects. Both numerator and
denominator are evaluated with the same fast-simulation card and the same
reconstruction-level analysis, so $C(m_T)$ isolates the response difference
driven by signal kinematics: any bias common to numerator and denominator —
an imperfect \textsc{Delphes} \cite{delphes} description of photon or jet reconstruction,
for instance — cancels to first order in the ratio, since it enters
$(N_{\rm reco}/N_{\rm fid})_{T\bar T}$ and
$(N_{\rm reco}/N_{\rm fid})_{t\bar t\gamma\gamma}^{\rm Delphes}$ in
essentially the same way. Consequently, the transfer factor isolates only
the response difference associated with the signal topology while
suppressing detector-modeling effects common to both processes.\\
The purpose of the transfer factor is to express the particle-level signal
prediction in the response convention of the published ATLAS fiducial
measurement. The effective signal contribution entering the likelihood is
therefore
\begin{equation}
\sigma_{\rm sig}^{\rm eff}(m_T,\mathbf{c})
= C(m_T)\,\sigma_{\rm sig}^{\rm fid}(m_T,\mathbf{c}),
\label{eq:atlaseffectivesignal_local}
\end{equation}
where the transfer factor is applied exclusively to the BSM signal
prediction. The measured cross section $\sigma_{\rm obs}$ and the published
SM prediction $\sigma_{\rm SM}$ in Eq.~\eqref{eq:atlasinputs} are left
untouched, since both already correspond to ATLAS's own fiducial and
reconstruction definitions and carry their own (uncorrected) uncertainties.\\
The SM validation sample consists of on-shell $t\bar t\gamma\gamma$
production with one semileptonic and one hadronic top decay, showered with
\textsc{Pythia} and passed through the ATLAS detector parameterisation in
\textsc{Delphes}. Particle-level jets are reconstructed independently with
\textsc{FastJet} using the same anti-$k_t$, $R=0.4$ definition as the
fiducial selection, and particle-level $b$ jets are identified by ghost
matching to $b$ hadrons with $p_T>5~\GeV$. The particle-level and
reconstruction-level selections applied to the SM sample are identical,
object for object, to those applied to every $T\bar T$ benchmark. This
ensures complete consistency between the validation sample and the signal
simulation, and is what makes the ratio in Eq.~\eqref{eq:correction}
meaningful.

\subsection{Closure test and associated systematic}

The fast detector simulation approximates the ATLAS detector response
but does not reproduce the complete reconstruction and
object-identification procedures of the ATLAS experiment. Its reconstruction efficiency must
therefore be validated against the SM reference quoted by ATLAS. Applying
the reconstruction-level and particle-level fiducial selections to the
SM validation sample gives
\begin{equation}
\epsilon_{\rm SM}^{\rm Delphes}
\equiv
\left(\frac{N_{\rm reco}}{N_{\rm fid}}\right)_{\rm SM}^{\rm Delphes}
=0.22,
\end{equation}
compared with the efficiency of $0.19$ quoted by ATLAS for the same fiducial
and reconstruction-level definitions. The relative non-closure between the
two is
\begin{equation}
\delta_{\rm closure}
=\frac{|0.22-0.19|}{0.19}\simeq 16\%.
\label{eq:atlasclosure}
\end{equation}
This discrepancy reflects residual differences between the public
\textsc{Delphes} \cite{delphes} parameterisation and the private ATLAS reconstruction
rather than a deficiency of the transfer-factor construction itself. 
We assign this discrepancy as a fully correlated, mass-independent
normalization uncertainty on the signal-response transfer factor.
The response uncertainty is propagated through the multiplicative
Gaussian-constrained nuisance parameter defined in
Eq.~\eqref{eq:atlasnuisancemorph}, with
$r_{\rm corr}=\delta_{\rm closure}$. The resulting relative
uncertainty on the signal prediction is propagated together with the
experimental uncertainties $\Delta\sigma_{\rm obs}$ and $\Delta\sigma_{\rm
SM}$ in the fiducial-cross-section likelihood, so that the imperfect
closure of the fast-simulation chain is reflected in the final exclusion
limits rather than absorbed silently into a point estimate of the transfer
factor.\\
For completeness, the same transfer factor can equivalently be expressed as
a corrected reconstruction efficiency for the signal,
\begin{equation}
\epsilon_{T\bar T}^{\rm corr}(m_T)
=\epsilon_{\rm SM}^{\rm ATLAS}\,C(m_T),
\qquad
\epsilon_{\rm SM}^{\rm ATLAS}=0.19,
\label{eq:atlascorrectedeff}
\end{equation}
which expresses the transfer factor in terms of an absolute signal reconstruction efficiency, using the SM efficiency quoted by ATLAS
as the reference normalization. This makes the size of the signal-dependent
response modification physically transparent. The two expressions are
equivalent: the statistical analysis applies $C(m_T)$ only once, through
Eq.~\eqref{eq:atlaseffectivesignal_local}, and does not multiply the
prediction by $\epsilon_{T\bar T}^{\rm corr}$ separately.\\
Table~\ref{tab:atlascorrectionnumbers} summarises the transfer factor
$C(m_T)$ and the corresponding corrected efficiency
$\epsilon_{T\bar T}^{\rm corr}$ across the simulated mass range. The
transfer factor decreases overall from $C(m_T)=1.63$ at $m_T=500~\GeV$ to
$C(m_T)=1.23$ at $m_T=3000~\GeV$. A small increase is observed between
$500$ and $750~\GeV$, after which the response decreases monotonically over
the remaining mass range, tracking the growing boost of the $T\bar T$ decay
products. As the final state becomes more collimated, the resolved-jet
multiplicity requirement, photon isolation, $b$-tagging, and the
jet--photon/jet--lepton overlap removal all become somewhat less efficient
at separating and correctly counting the individual objects, reducing the
response relative to its value at low mass. Expressed as an equivalent
corrected efficiency, this corresponds to a decrease from
$\epsilon_{T\bar T}^{\rm corr}\simeq0.31$--$0.32$ at low mass to
$0.23$ at $m_T=3~\TeV$. These results demonstrate that a single,
mass-independent response correction cannot accurately describe the
detector acceptance of the signal. A dedicated mass-dependent transfer
factor is therefore required for a faithful reinterpretation of the ATLAS
fiducial measurement.

\begin{table}[ht]
\centering
\caption{Mass-dependent detector-response transfer factor
$C(m_T)$ and the equivalent corrected reconstruction efficiency
$\epsilon_{T\bar T}^{\rm corr}=\epsilon_{\rm SM}^{\rm ATLAS}\,C(m_T)$, with
$\epsilon_{\rm SM}^{\rm ATLAS}=0.19$. The corrected efficiency is shown only
to illustrate the physical size of the response modification; the
statistical analysis applies solely the factor $C(m_T)$, through
Eq.~\eqref{eq:atlaseffectivesignal_local}.}
\label{tab:atlascorrectionnumbers}
\resizebox{\textwidth}{!}{%
\begin{tabular}{lccccccccccc}
\toprule
$m_T$ [GeV]
& 500 & 750 & 1000 & 1250 & 1500 & 1750 & 2000 & 2250 & 2500 & 2750 & 3000 \\
\midrule
$C(m_T)$
& 1.63 & 1.67 & 1.62 & 1.54 & 1.48 & 1.42 & 1.38 & 1.33 & 1.27 & 1.24 & 1.23 \\
$\epsilon_{T\bar T}^{\rm corr}$
& 0.31 & 0.32 & 0.31 & 0.29 & 0.28 & 0.27 & 0.26 & 0.25 & 0.24 & 0.24 & 0.23 \\
\bottomrule
\end{tabular}%
}
\end{table}

\section{EFT and narrow-width validity}
\label{sec:validity}

The transition interactions in Eq.~\eqref{eq:lagrangian} contain a
field-strength tensor and therefore introduce an explicit power of
momentum at each dipole vertex. Schematically, an amplitude containing
one dipole insertion scales as $cQ$, where $c$ denotes the corresponding
dimensionful coefficient and $Q$ is the characteristic momentum flowing
through the interaction. The truncation of the effective description is
meaningful only when this dimensionless expansion parameter remains
sufficiently small. At energies comparable to or larger than the scale
associated with the effective coefficient, contributions from additional
higher-dimensional interactions need no longer be suppressed relative
to the terms retained in the calculation
\cite{Manohar:2018aog,Brivio:2022pyi}.\\
This consideration is particularly relevant to the present analysis.
The enhanced population of energetic photons and heavy final states
provides sensitivity to the dipole interactions, but the same energetic
region is also where the effective description is most likely to become
unreliable. We therefore impose an explicit validity prescription at
every point in the coupling scan rather than quoting limits throughout
the full phenomenological parameter plane.\\
The cutoff associated with the effective interaction should not be
identified with the physical mass of the vector-like top partner. The
state $T$ is retained as an explicit propagating degree of freedom and is
produced on shell, whereas the dipole interactions parametrize additional
dynamics that has been integrated out. In a specified ultraviolet
completion, the relation between the effective coefficients and the
scale of the omitted states would generally depend on mediator masses,
couplings, mixing angles, loop factors, and the normalization of the
underlying gauge-invariant operators. Because the transition interactions are parametrized directly in terms of
physical fields after electroweak symmetry breaking, the two neutral
transition coefficients alone do not uniquely determine the mass scale of
the underlying ultraviolet physics.
For the purpose of defining a transparent and reproducible validity
criterion, we introduce the point-dependent effective scale
\begin{equation}
\Lambda_{\rm eff}(\boldsymbol{c})
=
\frac{1}{\sqrt{c_{tg}^{\,2}+c_{t\gamma}^{\,2}}},
\qquad
\boldsymbol{c}=(c_{t\gamma},c_{tg}).
\label{eq:lambdaeff}
\end{equation}
This definition is a phenomenological validity convention for the two
scanned neutral transition coefficients. It neither restores electroweak
gauge invariance nor identifies $\Lambda_{\rm eff}$ with the mass of a
specific ultraviolet mediator.

\subsection{Event-level EFT-validity requirement}
\label{sec:eftenergy}

We use the invariant mass of the produced heavy-quark pair,
$M_{T\bar T}$, as the hard scale characterizing the production
subprocess. This quantity measures the total partonic center-of-mass
energy entering the production of the heavy pair, is obtained directly
from the hard-process event record, and is independent of the subsequent
reconstruction of the $T$-quark decay products. Although the momentum
transfer through an individual dipole vertex can be smaller than
$M_{T\bar T}$ in some regions of phase space, the choice
$Q=M_{T\bar T}$ provides a transparent and conservative measure of the
energy probed by the hard scattering.\\
For every tested point
$\boldsymbol{c}=(c_{t\gamma},c_{tg})$ in the coupling plane,
$\Lambda_{\rm eff}(\boldsymbol{c})$ is recalculated using
Eq.~\eqref{eq:lambdaeff}. An event contribution is retained only when
\begin{equation}
M_{T\bar T}<\Lambda_{\rm eff}(\boldsymbol{c}).
\label{eq:truncation}
\end{equation}
This requirement is imposed event by event, separately in every analysis
bin and for each EFT-dependent production component. Schematically, the
truncated contribution to bin $i$ is
\begin{equation}
\sigma_{i,X}^{\rm trunc}(\boldsymbol{c})
=
\sum_{e\in i}
w_{e,X}(\boldsymbol{c})\,
\Theta\!\left[
\Lambda_{\rm eff}(\boldsymbol{c})-M_{T\bar T,e}
\right],
\qquad
X\in\{\mathrm{int},\mathrm{dip}\},
\label{eq:weightedtrunc}
\end{equation}
where $w_{e,X}(\boldsymbol{c})$ is the properly normalized,
coupling-dependent weight of event $e$ for production component $X$.
The label $X=\mathrm{int}$ denotes the interference between the
pure-QCD and dipole-induced amplitudes, while $X=\mathrm{dip}$ denotes
the squared dipole-induced contribution. The step function retains
events satisfying Eq.~\eqref{eq:truncation} and removes those at or above
the effective cutoff.
The interference contribution is not a separately positive cross
section and can contain both positive and negative event weights.
Signed weights are therefore preserved when applying
Eq.~\eqref{eq:weightedtrunc}. The truncation is performed on the
individual weighted events before summing them within an analysis bin.
This procedure preserves the cancellations dictated by the matrix
element within the retained phase-space region. Parameter points for
which strong cancellations lead to an insufficient effective Monte
Carlo sample size are regenerated or omitted rather than stabilized
through an uncontrolled interpolation.
Both the event weights and the cutoff depend on the tested coefficients.
Consequently, the set of retained events changes across the
$(c_{t\gamma},c_{tg})$ plane, and the truncation can modify both the
normalization and the kinematic shape of the signal. The truncated
templates are therefore reconstructed at every scan point before the
likelihood defined in Sec.~\ref{sec:statistics} is evaluated. They are
not obtained by multiplying the untruncated prediction by a single
inclusive survival factor.\\
The pure-QCD $T\bar T$ production amplitude contains no
higher-dimensional production vertex and is therefore not subjected to
the production-level truncation in Eq.~\eqref{eq:truncation}. The
radiative decays $T\to t\gamma$ and $T\to tg$, however, are themselves
mediated by the effective transition interactions. The interpretation
of the pure-QCD production contribution followed by a radiative decay
therefore additionally requires
\begin{equation}
m_T<\Lambda_{\rm eff}(\boldsymbol{c}).
\label{eq:decayeftvalidity}
\end{equation}
For an EFT-dependent production event satisfying
Eq.~\eqref{eq:truncation}, this decay-level condition follows
automatically from the on-shell pair-production threshold
\begin{equation}
M_{T\bar T}\geq 2m_T.
\label{eq:pairthreshold}
\end{equation}
It must nevertheless be imposed explicitly on the pure-QCD production
component, which is retained independently of the production-level
truncation.
Equation~\eqref{eq:pairthreshold} also provides a direct validation of
the numerical implementation. If
$\Lambda_{\rm eff}\leq2m_T$, no on-shell $T\bar T$ event can satisfy
Eq.~\eqref{eq:truncation}. The truncated interference and
dipole-squared production contributions must therefore vanish exactly.
This behavior was verified in the generated samples. \\
The numerical impact of the event-level truncation can be illustrated at a
few representative points. For $m_T=500~\GeV$, the fraction of the
EFT-dependent fiducial contribution satisfying
$M_{T\bar T}<\Lambda_{\rm eff}$ is approximately $0.70$ in the CMS
$t\bar t\gamma$ analysis and $0.80$ in the ATLAS
$t\bar t\gamma\gamma$ analysis for
$\Lambda_{\rm eff}=1500~\GeV$. Increasing the cutoff to
$\Lambda_{\rm eff}=2000~\GeV$ raises these fractions to approximately
$0.95$ and $0.96$, respectively. As a higher-mass example, for
$m_T=1250~\GeV$ and $\Lambda_{\rm eff}=4000~\GeV$, the corresponding
survival fractions are approximately $0.97$ for CMS and $0.95$ for
ATLAS. This behavior follows directly from the kinematics of heavy-quark pair
production. For an on-shell $T\bar T$ system,
$M_{T\bar T}\geq2m_T$, and therefore no EFT-dependent event can satisfy
the truncation requirement when $\Lambda_{\rm eff}\leq2m_T$. Above this
threshold, the surviving fraction increases with $\Lambda_{\rm eff}$ and
approaches unity when the cutoff lies well above the invariant masses
populated by the selected events. Although $M_{T\bar T}$ is determined from the production-level event
record, the fiducial selections can alter its distribution among accepted
events. Differences between the CMS and ATLAS survival efficiencies may
therefore reflect their different selections as well as finite-simulation
statistical fluctuations.\\
A valid exclusion is obtained up to $m_T=1250~\GeV$. For
$m_T\geq1500~\GeV$, however, no statistically excluded region remains
after the EFT-validity conditions are imposed. This loss of valid
sensitivity results from the competing dependence on the transition
couplings. Coupling values large enough to produce an experimentally
resolvable contribution correspond to a smaller
$\Lambda_{\rm eff}=1/\sqrt{c_{t\gamma}^2+c_{tg}^2}$ and hence remove an
increasing fraction of the high-$M_{T\bar T}$ signal. Conversely, reducing
the couplings raises the cutoff and improves the EFT survival efficiency,
but simultaneously suppresses the dipole-induced production rate below
the sensitivity of the Run~2 measurements. The rapid decrease of the
$T\bar T$ production cross section with increasing $m_T$ further
strengthens this tension. Consequently, the untruncated analysis can show
nominal sensitivity at masses of $1500~\GeV$ and above, but that sensitivity
does not overlap with the domain in which the effective description is
self-consistent. No EFT-valid limits are therefore reported for these
higher masses.\\
The quoted survival fractions are illustrative rather than universal
correction factors. In the statistical analysis, $\Lambda_{\rm eff}$ is
recalculated at every $(c_{t\gamma},c_{tg})$ point, and the EFT-dependent
contribution is truncated event by event before the corresponding CMS or
ATLAS likelihood is evaluated.

\subsection{Narrow-width requirement}
\label{sec:widthvalidity}

The event generation factorizes on-shell $T\bar T$ production from the
subsequent heavy-quark decays. The physical interpretation therefore
requires the narrow-width approximation to remain reliable. At every
coupling point, the complete width
$\Gamma_T$
is evaluated using the same coefficients and model parameters as those
used for the signal prediction. Any additional decay mode present in the
implemented model must likewise be included if its partial width is
nonzero. We retain only points satisfying
$\Gamma_T(c_{tg},c_{t\gamma},m_T)/m_T<0.1.$
This requirement is imposed because the signal samples
are generated in the narrow-width approximation, in which the production
of the on-shell $T\bar T$ pair is separated from the subsequent decays of
the heavy quarks. This factorized description is reliable only when the
width of $T$ is small compared with its mass. For a larger width, the
intermediate $T$ quark can be produced farther from its mass shell, which
changes the invariant-mass distribution and may also modify the fraction
of signal events passing the fiducial selection. In addition, interference
with nonresonant amplitudes leading to the same final state can become
important. These finite-width and interference effects are not included
in the on-shell samples and cannot be accounted for simply by changing
the total width or rescaling the signal with different branching fractions
\cite{Deandrea:2021vje}.
The EFT-energy and narrow-width requirements test different aspects of
the calculation and are therefore imposed independently. A point can
have a sufficiently narrow resonance while failing the EFT-energy
condition, or it can satisfy the EFT-energy requirement while failing
the width criterion because the dipole-induced partial widths grow
quadratically with the effective coefficients. A parameter point is
reported as physically interpretable only if it satisfies both
requirements.
The event-level EFT truncation is applied when constructing the signal
templates and therefore modifies the likelihood prediction at each
coupling point. By contrast, the narrow-width condition 
is applied as an interpretive mask to the
resulting likelihood scan. The contour
$\Gamma_T/m_T=0.1$ is overlaid on the coupling-plane figures, and the
region with $\Gamma_T/m_T\geq0.1$ is not reported as part of the final
physical exclusion.

\section{Statistical analysis}
\label{sec:statistics}

The CMS and ATLAS measurements are interpreted independently. No
statistical combination is performed because the information required to
model correlations between the two measurements is not publicly
available. For each value of $m_T$, the parameters of interest are the
transition coefficients
$\boldsymbol{c}=(c_{t\gamma},c_{tg})$. The physical branching fractions and signal normalizations are
recalculated at every tested point in this coupling plane.

\subsection{CMS differential likelihood}
\label{sec:cmslikelihood}

The CMS reinterpretation uses only the absolute particle-level
photon-transverse-momentum distribution
$d\sigma/dp_T^\gamma$. The measured cross sections are collected in the
vector $\boldsymbol{d}$, while $\boldsymbol{b}$ and
$\boldsymbol{s}$ denote the corresponding SM and heavy-$T$ predictions,
respectively. All three vectors use the same photon-$p_T$ binning and overflow-bin
convention as the CMS measurement; only the last four bins are included
in the fit.\\
The signal vector includes both radiative topologies that can satisfy the
exactly-one-photon fiducial requirement:
\begin{equation}
\boldsymbol{s}(m_T,\boldsymbol{c})
=
\boldsymbol{s}_{\gamma g}^{(1\gamma_{\rm fid})}
(m_T,\boldsymbol{c})
+
\boldsymbol{s}_{\gamma\gamma}^{(1\gamma_{\rm fid})}
(m_T,\boldsymbol{c}),
\label{eq:cmssignalvector}
\end{equation}
where
$\boldsymbol{c}=(c_{t\gamma},c_{tg})$. The mixed and
double-radiative contributions are evaluated independently in every
$p_T^\gamma$ bin and normalized using the physical branching-fraction
factors $2\mathcal{B}_{\gamma}\mathcal{B}_{g}$ and
$\mathcal{B}_{\gamma}^{2}$, respectively.
The CMS likelihood is written as
\begin{align}
-2\ln L_{\rm CMS}(\boldsymbol{c},\boldsymbol{\theta})
={}&
\left[
\boldsymbol{d}
-\boldsymbol{b}(\boldsymbol{\theta})
-\boldsymbol{s}(m_T,\boldsymbol{c},\boldsymbol{\theta})
\right]^{T}
V_{\rm CMS}^{-1}
\left[
\boldsymbol{d}
-\boldsymbol{b}(\boldsymbol{\theta})
-\boldsymbol{s}(m_T,\boldsymbol{c},\boldsymbol{\theta})
\right]
\nonumber\\
&+\sum_k\theta_k^2 ,
\label{eq:cmslikelihood}
\end{align}
where $\boldsymbol{\theta}$ denotes nuisance parameters associated with
uncertainty sources not already included in the published experimental
covariance matrix.
The experimental inputs are taken from version~3 of the HEPData record
associated with the CMS measurement
\cite{CMS:2022lmh,CMS:HEPData:ttgamma}. The absolute photon-$p_T$
cross sections and their statistical and systematic uncertainties are
taken from the CMS paper \cite{CMS:2022lmh,CMS:HEPData:ttgamma}.
The statistical and systematic correlation matrices are taken from the
associated supplementary HEPData tables.
The HEPData values correspond to absolute cross sections integrated over
each photon-$p_T$ bin and are not divided by the bin width. To construct
the differential distribution used here, the cross section and its
statistical and systematic uncertainties in bin $i$ are divided by the
corresponding bin width $\Delta p_{T,i}^{\gamma}$. The same bin-width
transformation is applied to the measured data, SM prediction, signal
prediction, and covariance matrix.
The experimental covariance matrix is constructed as
\begin{equation}
\left(V_{\rm CMS}^{\rm bin}\right)_{ij}
=
\rho_{ij}^{\rm stat}\,
\Delta_{i,\rm bin}^{\rm stat}
\Delta_{j,\rm bin}^{\rm stat}
+
\rho_{ij}^{\rm syst}\,
\Delta_{i,\rm bin}^{\rm syst}
\Delta_{j,\rm bin}^{\rm syst},
\label{eq:cmscovariancebin}
\end{equation}
where $\rho_{ij}^{\rm stat}$ and $\rho_{ij}^{\rm syst}$ are the
statistical and systematic correlation coefficients provided by
HEPData, and $\Delta_{i,\rm bin}^{\rm stat}$ and
$\Delta_{i,\rm bin}^{\rm syst}$ are the corresponding absolute
uncertainties on the bin-integrated cross section.
Defining the diagonal bin-width matrix
\begin{equation}
D_{ij}
=
\frac{\delta_{ij}}{\Delta p_{T,i}^{\gamma}},
\end{equation}
the covariance matrix for the differential cross sections entering
Eq.~\eqref{eq:cmslikelihood} is
\begin{equation}
V_{\rm CMS}
=
D\,V_{\rm CMS}^{\rm bin}D^{T}.
\label{eq:cmscovariance}
\end{equation}
This transformation preserves the published statistical and systematic
correlations while converting the covariance consistently to the units
of $d\sigma/dp_T^\gamma$.
Only the absolute $p_T^\gamma$ distribution is used in the likelihood.
No normalized distribution and no second CMS observable are included.
Consequently, no covariance between different observables is assumed or
required. An experimental uncertainty already represented in
$V_{\rm CMS}$ is not introduced again through an additional nuisance
parameter.
Systematic variations of a positive signal component are interpolated
continuously between the nominal upward and downward templates. PDF
variations derived from common event weights are treated as correlated
among the affected $p_T^\gamma$ bins. Renormalization- and
factorization-scale variations are retained as separate uncertainty
sources. Finite-simulation uncertainties are treated as bin-local unless
an explicit correlation arises from a common weighted event sample.
The QCD correction factor is applied only to the pure-QCD
$T\bar T$ production component. It is not applied to the
QCD--dipole interference, the dipole-squared production contribution,
the decay widths, or the branching fractions.

\subsection{ATLAS counting likelihood and confidence regions}
\label{sec:atlaslikelihood}

The ATLAS $t\bar t\gamma\gamma$ measurement is interpreted as a single
fiducial counting observable. The measured cross section and SM prediction
used in the analysis are
\begin{equation}
\sigma_{\rm obs}=2.51~\fb,
\qquad
\Delta\sigma_{\rm obs}=0.60~\fb,
\qquad
\sigma_{\rm SM}=1.53~\fb,
\qquad
\Delta\sigma_{\rm SM}=0.46~\fb .
\label{eq:atlasstatinputs}
\end{equation}
The effective BSM contribution is
$\sigma_{\rm sig}^{\rm eff}
=C(m_T)\sigma_{\rm sig}^{\rm fid}$, with the transfer factor
$C(m_T)$ defined in Sec.~\ref{sec:atlas} and applied once to the signal.\\
PDF, factorization-scale, renormalization-scale, and response uncertainties
are represented by independent, Gaussian-constrained nuisance parameters
$\theta_k$. Their multiplicative variations are
\begin{equation}
f_k(\theta_k)
=
\exp\!\left[\ln(1+r_k)\,\theta_k\right],
\label{eq:atlasnuisancemorph}
\end{equation}
where $k\in\{\mathrm{PDF},\mu_F,\mu_R,\mathrm{corr}\}$ and
$r_k$ denotes the corresponding relative uncertainty.
The PDF and scale values are listed in
Table~\ref{tab:atlassignalunc}, while
$r_{\rm corr}\equiv\delta_{\rm closure}=0.16$ is the
response-closure uncertainty defined in Eq.~\eqref{eq:atlasclosure}.\\
The ATLAS test statistic is
\begin{equation}
\chi^2_{\rm ATLAS}(m_T,\mathbf c,\boldsymbol\theta)
=
\frac{
\left[
\sigma_{\rm obs}-\sigma_{\rm SM}
-\sigma_{\rm sig}^{\rm eff}(m_T,\mathbf c)
\prod_k f_k(\theta_k)
\right]^2
}{
(\Delta\sigma_{\rm obs})^2+(\Delta\sigma_{\rm SM})^2
}
+\sum_k\theta_k^2 .
\label{eq:atlaschi2}
\end{equation}
The experimental and SM uncertainties are treated as independent.
Signal uncertainties enter only through the nuisance parameters and
are not added separately to the denominator.

\begin{table}[ht]
\centering
\caption{Dimensionless relative PDF, factorization-scale, and
renormalization-scale uncertainties used to define the ATLAS
$t\bar t\gamma\gamma$ signal nuisance parameters through
Eq.~\eqref{eq:atlasnuisancemorph}. For example, $0.026$ corresponds to
$2.6\%$. The additional relative response uncertainty is
$\delta_{\rm closure}=0.16$ at every mass.}
\label{tab:atlassignalunc}
\resizebox{\textwidth}{!}{%
\begin{tabular}{lrrrrrrrrrrr}
\toprule
$m_T$ [GeV]
& 500 & 750 & 1000 & 1250 & 1500 & 1750
& 2000 & 2250 & 2500 & 2750 & 3000 \\
\midrule
$r_{\rm PDF}$
& 0.026 & 0.037 & 0.048 & 0.062 & 0.087 & 0.13
& 0.21 & 0.30 & 0.45 & 0.57 & 0.69 \\
$r_{\mu_F}$
& 0.11 & 0.13 & 0.14 & 0.15 & 0.15 & 0.16
& 0.17 & 0.17 & 0.17 & 0.18 & 0.18 \\
$r_{\mu_R}$
& 0.34 & 0.33 & 0.32 & 0.31 & 0.30 & 0.29
& 0.29 & 0.29 & 0.29 & 0.29 & 0.28 \\
\bottomrule
\end{tabular}}
\end{table}

For each analysis separately and at fixed $m_T$, nuisance parameters
are profiled and the confidence-region statistic is defined as
$\Delta\chi^2(m_T,\mathbf c)=\chi^2_{\rm prof}(m_T,\mathbf c)-\chi^2_{\min}(m_T)$, where $\chi^2_{\min}$ is the global minimum of the same profiled
statistic over the coupling plane. 

\section{Results}
\label{sec:results}

The constraints are presented in the plane of the two transition
coefficients $(c_{t\gamma},c_{tg})$ defined in
Eq.~\eqref{eq:effectivecouplings}. The CMS and ATLAS measurements are interpreted
separately, and no statistical combination is performed. The Run~2 contours
represent observed 95\% confidence-level constraints, whereas the HL-LHC
contours represent projected 95\% confidence-level sensitivities for two
parameters of interest. All signal predictions incorporate the
point-dependent, event-level EFT truncation defined in
Sec.~\ref{sec:validity}. Only the portions of the resulting exclusions that
also satisfy $\Gamma_T/m_T<0.1$ are retained in the physical
interpretation.

\subsection{CMS $t\bar t\gamma$ reinterpretation}
\label{sec:cmsresults}

\begin{figure}[!ht]
\centering
\includegraphics[width=0.72\textwidth]
{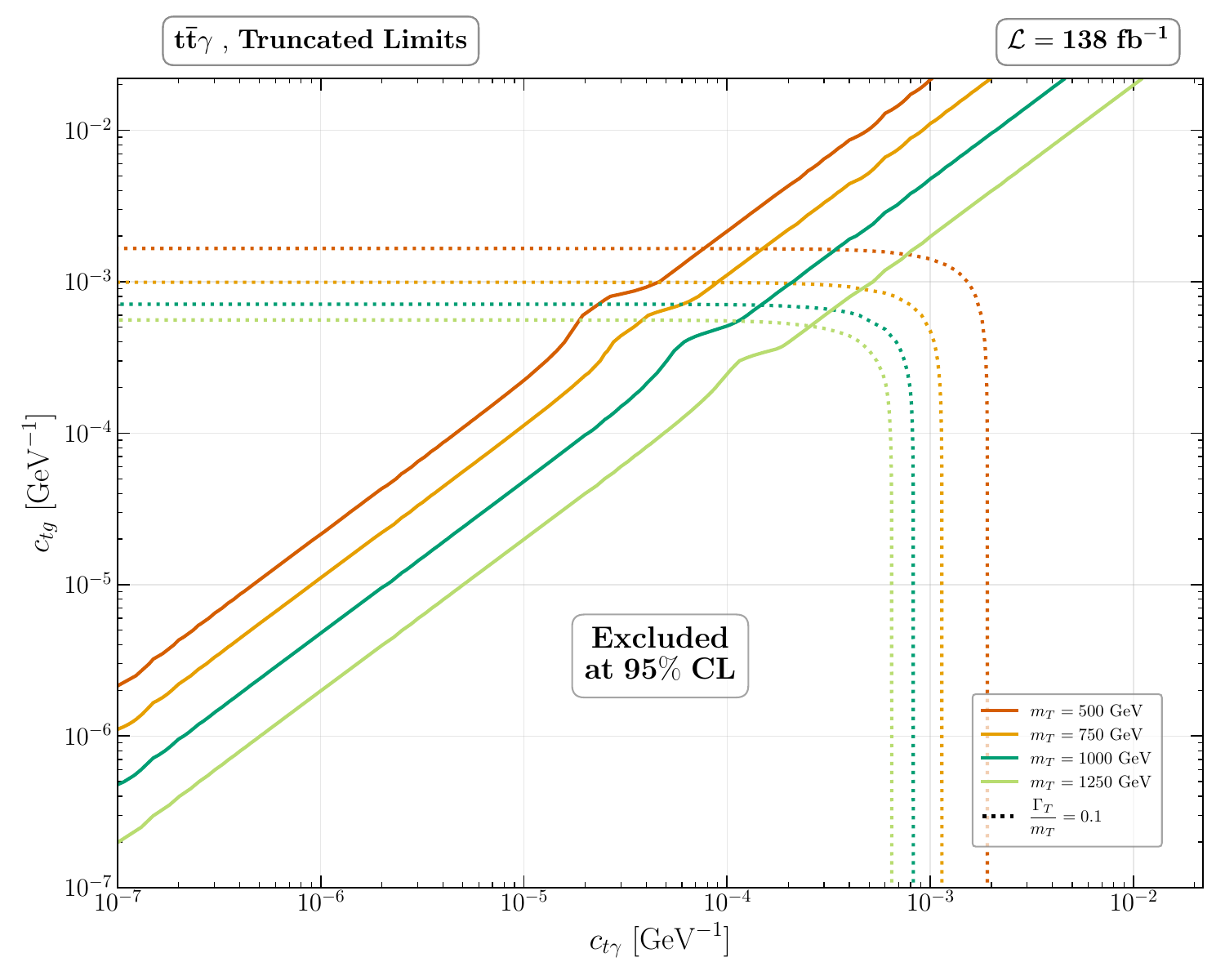}
\caption{Observed 95\% CL constraints in the
$(c_{t\gamma},c_{tg})$ plane from the CMS $t\bar t\gamma$
reinterpretation using the Run~2 data set corresponding to
$138~\fb^{-1}$. The solid colored curves are obtained after applying the
coupling-dependent EFT truncation event by event and bin by bin. Each color
denotes a different $T$-quark mass. The dotted curves show the corresponding
$\Gamma_T/m_T=0.1$ boundaries, and the excluded side of the statistical
contours is indicated in the figure. Only the portions of the statistical
exclusions satisfying both the EFT-energy and narrow-width requirements are
retained in the physical interpretation.}
\label{fig:cmsrun2}
\end{figure}

The CMS reinterpretation uses the absolute particle-level
$d\sigma/dp_T^\gamma$ spectrum. The signal in each bin contains the mixed
topology
\begin{equation}
T\bar T\to(t\gamma)(\bar t g),
\end{equation}
including its charge-conjugate assignment, and the contribution from
\begin{equation}
T\bar T\to(t\gamma)(\bar t\gamma)
\end{equation}
events that satisfy the exactly-one-fiducial-photon requirement. The two
contributions are evaluated independently and normalized using the physical
branching fractions. The likelihood uses the full bin-to-bin covariance
matrix and the signal-uncertainty treatment described in
Sec.~\ref{sec:statistics}.\\
The observed Run~2 constraints are shown in
Fig.~\ref{fig:cmsrun2}. The solid colored curves are the 95\% CL boundaries
obtained using the event-level-truncated signal templates, while the dotted
curves denote $\Gamma_T/m_T=0.1$ for each mass hypothesis. The excluded side
of the statistical contours is indicated in the figure. Only the overlap of
the statistical exclusion with the EFT- and width-valid domain is retained in
the physical interpretation.\\
For a quantitative characterization of the two-dimensional result, we
consider two nonzero benchmark slices. At fixed
$c_{tg}=1.0\times10^{-5}~\GeV^{-1}$, the observed Run~2 exclusions
extend down to
\begin{equation}
c_{t\gamma}\simeq
\left(
4.7\times10^{-7},\,
9.1\times10^{-7},\,
2.1\times10^{-6},\,
5.1\times10^{-6}
\right)~\GeV^{-1}
\label{eq:cmsrun2ctgamma}
\end{equation}
for $m_T=500$, $750$, $1000$, and $1250~\GeV$, respectively.
The excluded intervals extend from these values toward larger
$c_{t\gamma}$ until the corresponding narrow-width boundary,
$\Gamma_T/m_T=0.1$, is reached.\\
Along the complementary slice
$c_{t\gamma}=1.0\times10^{-5}~\GeV^{-1}$, the observed Run~2 exclusions
extend up to
\begin{equation}
c_{tg}\simeq
\left(
2.2\times10^{-4},\,
1.1\times10^{-4},\,
4.8\times10^{-5},\,
2.0\times10^{-5}
\right)~\GeV^{-1}.
\label{eq:cmsrun2ctg}
\end{equation}
For this benchmark slice, the region below the relevant contour boundary
is excluded within the narrow-width domain, $\Gamma_T/m_T<0.1$.\\
The distinct behavior along the two slices reflects the topology dependence
of the CMS signal. The mixed-radiative contribution is proportional to
$2\mathcal B_\gamma\mathcal B_g$ and is largest when neither radiative
branching fraction is strongly suppressed. The migrated double-radiative
contribution is instead proportional to $\mathcal B_\gamma^2$ and preserves
sensitivity as $c_{tg}$ becomes small. The contours must consequently be
interpreted as genuinely two-dimensional constraints: neither coefficient
has a unique one-dimensional limit unless the value of the other coefficient,
or an equivalent relation between them, is specified.

\subsection{ATLAS $t\bar t\gamma\gamma$ reinterpretation}
\label{sec:atlasresults}

\begin{figure}[!ht]
\centering
\includegraphics[width=0.72\textwidth]
{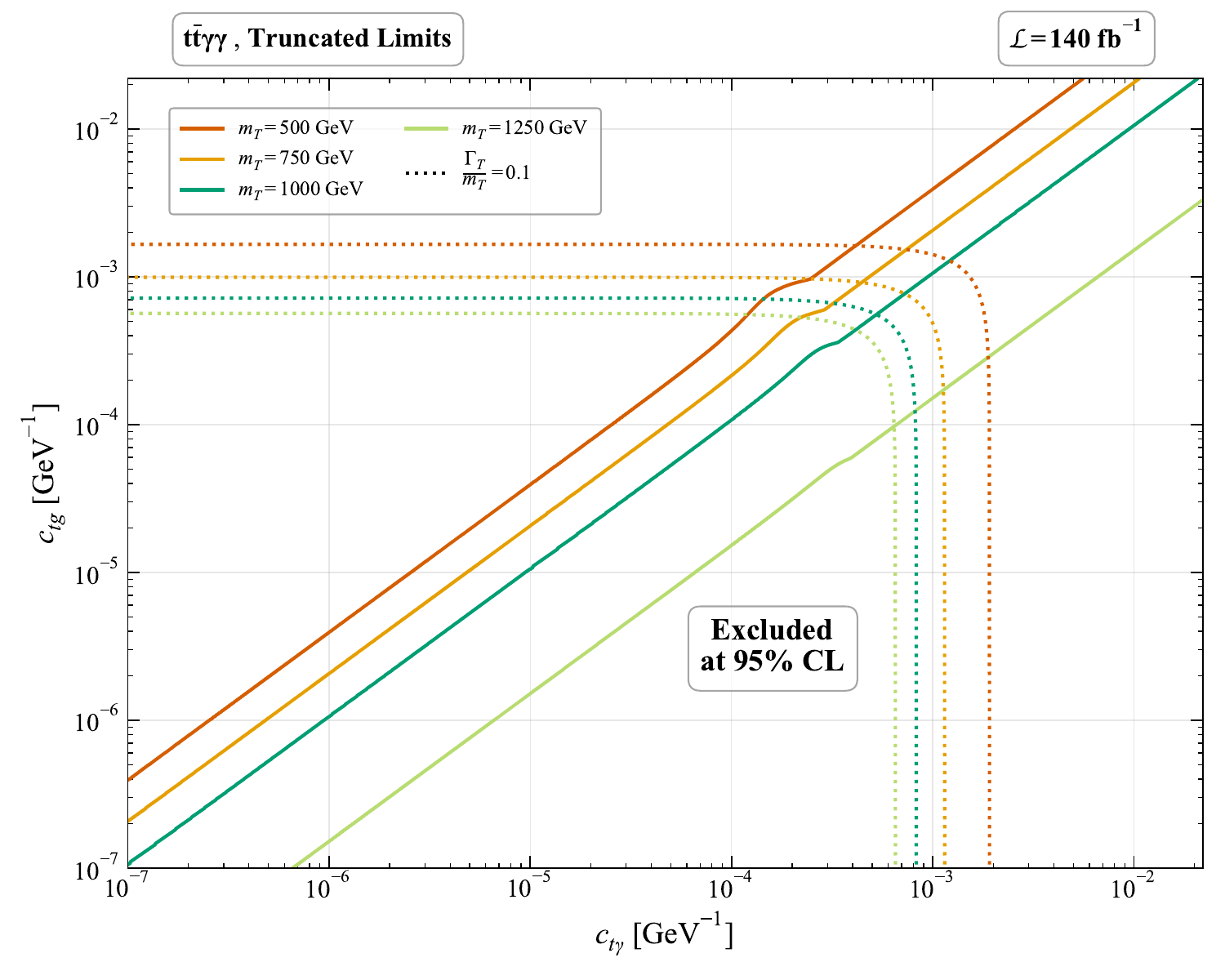}
\caption{Observed 95\% CL constraints in the
$(c_{t\gamma},c_{tg})$ plane from the ATLAS
$t\bar t\gamma\gamma$ counting reinterpretation using the Run~2 data set
corresponding to $140~\fb^{-1}$. The solid colored curves are recomputed
after applying the point-dependent EFT truncation event by event. Each color
represents a different $T$-quark mass. The dotted curves denote the
corresponding $\Gamma_T/m_T=0.1$ boundaries, and the excluded side of the
statistical contours is indicated in the figure. Only the portions of the
statistical exclusions satisfying both the EFT-energy and narrow-width
requirements are retained as physical constraints.}
\label{fig:atlasrun2}
\end{figure}

The ATLAS measurement is interpreted as a single fiducial counting
observable.
The mass-dependent detector-response transfer factor $C(m_T)$, its
16\% closure uncertainty, and the relative PDF and scale uncertainties
are incorporated as described in Secs.~\ref{sec:atlas} and
\ref{sec:statistics}.
The selected signal is the double-radiative topology
$T\bar T\to(t\gamma)(\bar t\gamma)$, whose normalization is proportional
to $\mathcal B_\gamma^2$. The ATLAS measurement is therefore particularly
sensitive to regions in which the electromagnetic transition is not strongly
suppressed relative to the chromomagnetic transition.\\
The observed Run~2 constraints are presented in
Fig.~\ref{fig:atlasrun2}. At fixed
$c_{tg}=1.0\times10^{-5}~\GeV^{-1}$, the exclusions extend down to
\begin{equation}
c_{t\gamma}\simeq
\left(
2.3\times10^{-6},\,
4.4\times10^{-6},\,
8.5\times10^{-6},\,
5.8\times10^{-5}
\right)~\GeV^{-1}
\label{eq:atlasrun2ctgamma}
\end{equation}
for $m_T=500$, $750$, $1000$, and $1250~\GeV$, respectively.
At fixed
$c_{t\gamma}=1.0\times10^{-5}~\GeV^{-1}$, the exclusions extend up to
\begin{equation}
c_{tg}\simeq
\left(
4.2\times10^{-5},\,
2.3\times10^{-5},\,
1.2\times10^{-5},\,
1.8\times10^{-6}
\right)~\GeV^{-1}.
\label{eq:atlasrun2ctg}
\end{equation}
These exclusion reaches are quoted within the narrow-width domain,
$\Gamma_T/m_T<0.1$.\\
The pronounced dependence on the coupling ratio is a direct consequence of
the double-radiative branching fraction. Increasing $c_{tg}$ at fixed
$c_{t\gamma}$ enhances the gluonic partial width and suppresses
$\mathcal B_\gamma$, thereby reducing the
$t\bar t\gamma\gamma$ signal rate. The loss of sensitivity is amplified at
larger $m_T$ by the rapidly falling $T\bar T$ production cross section.
The ATLAS contours therefore constrain a correlated region of the
two-coupling plane rather than providing independent limits on
$c_{t\gamma}$ and $c_{tg}$.

For both reinterpretations, EFT- and width-valid sensitivity is retained
through $m_T=1250~\GeV$, whereas no valid exclusion remains for
$m_T\geq1500~\GeV$. This loss of high-mass reach reflects the competition
between experimental sensitivity and EFT validity. The QCD
$T\bar T$ production cross section decreases rapidly with increasing
$m_T$, so larger transition coefficients are required to produce an
observable signal. Larger coefficients, however, correspond to a smaller
$\Lambda_{\rm eff}$ and cause a greater fraction of the energetic
EFT-dependent contribution to fail
$M_{T\bar T}<\Lambda_{\rm eff}$. Conversely, smaller coefficients improve
the validity of the effective description but yield insufficient sensitivity.
At high masses, the statistically excluded and theoretically valid regions
therefore cease to overlap.\\
The larger HL-LHC data set partly reduces this tension by extending the
sensitivity toward smaller transition coefficients and hence larger values of
$\Lambda_{\rm eff}$. The projected contours follow the uncertainty-scaling
assumptions specified in Sec.~\ref{sec:hllhcresults} and are not complete
forecasts of future detector performance or theoretical precision.

\subsection{HL-LHC projections}
\label{sec:hllhcresults}

\begin{figure}[!ht]
\centering
\includegraphics[width=0.49\textwidth]
{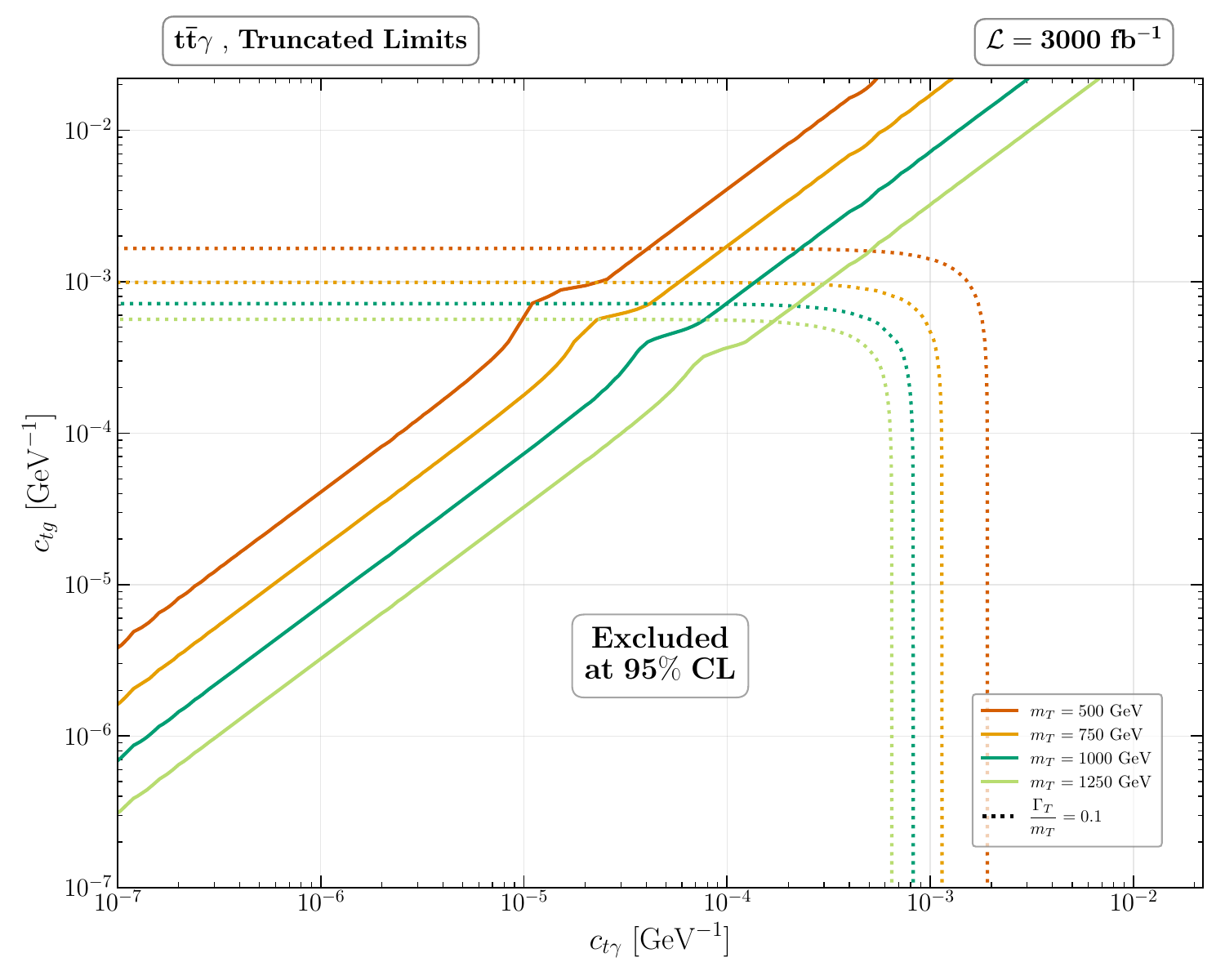}
\hfill
\includegraphics[width=0.49\textwidth]
{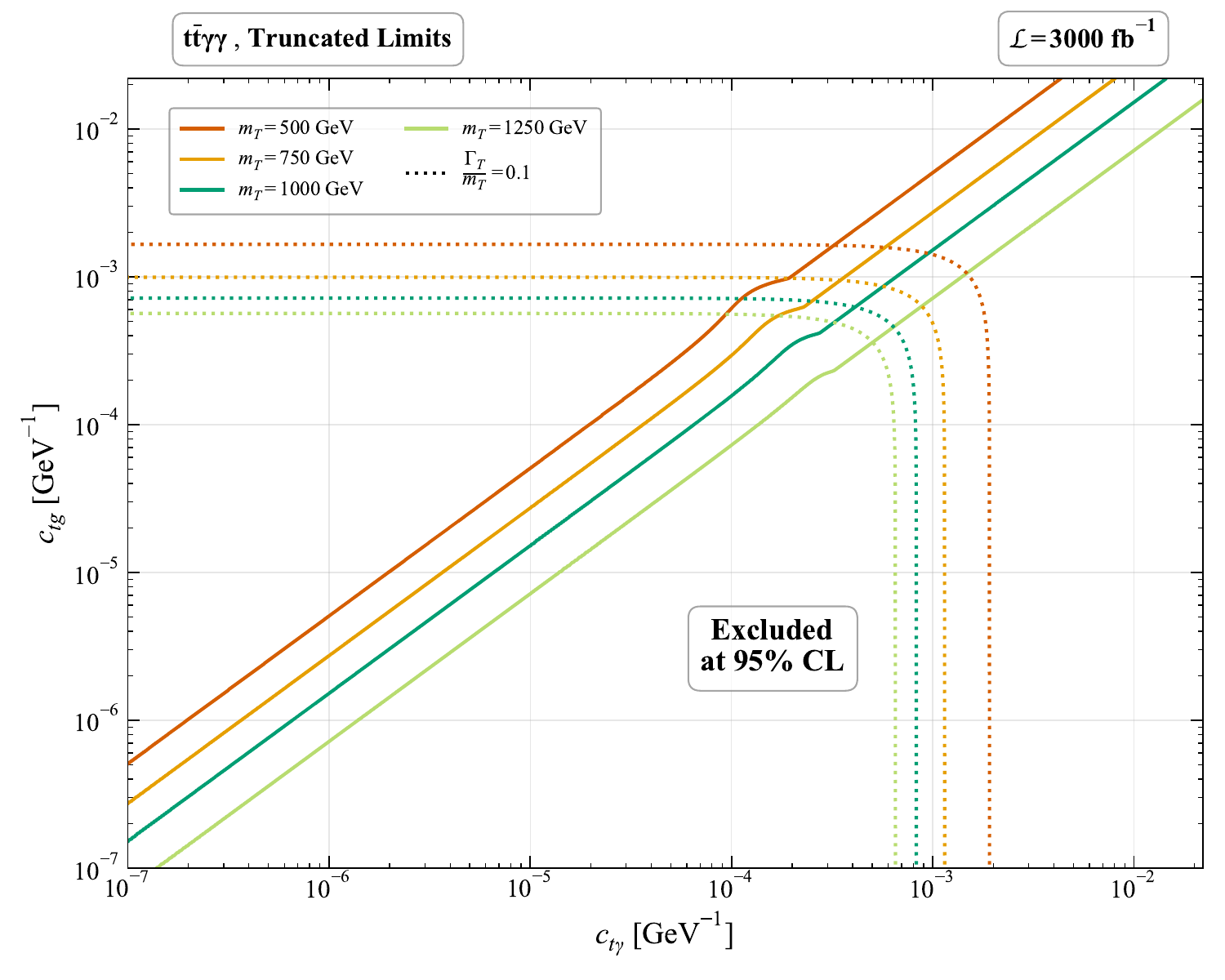}
\caption{Projected 95\% CL sensitivities in the
$(c_{t\gamma},c_{tg})$ plane at the HL-LHC for an integrated luminosity
of $3000~\fb^{-1}$. The left panel shows the projection based on the CMS
$t\bar t\gamma$ differential reinterpretation, and the right panel shows
the projection based on the ATLAS $t\bar t\gamma\gamma$ counting
reinterpretation. The solid colored curves are calculated after applying the
coupling-dependent EFT truncation event by event; for CMS, the truncation is
also implemented independently in every $p_T^\gamma$ bin. Each color denotes
a different $T$-quark mass. The dotted curves show the corresponding
$\Gamma_T/m_T=0.1$ boundaries. The physical interpretation is restricted
to the portions of the EFT-truncated statistical exclusions satisfying
$\Gamma_T/m_T<0.1$. For CMS, the statistical uncertainty is scaled by
$\sqrt{138/3000}$ and the Run~2 experimental systematic uncertainties
are halved. For ATLAS, the total reported experimental uncertainty is
scaled by $\sqrt{140/3000}$. The SM-theory, signal-theory,
finite-simulation, migration, and response uncertainties remain unchanged
in both analyses.}
\label{fig:hllhcconstraints}
\end{figure}

The CMS and ATLAS sensitivities are projected to an integrated luminosity of
$3000~\fb^{-1}$. The resulting contours are presented in
Fig.~\ref{fig:hllhcconstraints}, with the CMS projection in the left panel
and the ATLAS projection in the right panel. For CMS, the statistical
component of the Run~2 experimental uncertainty is rescaled as
\begin{equation}
\Delta_{\rm stat}^{\rm HL}
=
\Delta_{\rm stat}^{\rm Run\,2}
\sqrt{\frac{138~\fb^{-1}}
{3000~\fb^{-1}}},
\label{eq:hllhcstatscaling}
\end{equation}
while the Run~2 experimental systematic uncertainties are reduced
by a factor of two.\\
For the ATLAS $t\bar t\gamma\gamma$ channel, the statistical and
systematic components of the reported experimental uncertainty are not
available separately in the inputs used here. We therefore adopt a
luminosity-scaled total-experimental-uncertainty scenario, in which the
total reported experimental uncertainty is multiplied by
$\sqrt{140/3000}$. This prescription assumes that the total experimental
uncertainty decreases as $1/\sqrt{\mathcal L}$; it does not represent a
statistical-only extrapolation and may overestimate the improvement if
systematic uncertainties remain irreducible.\\
For both analyses, the SM-theory, signal-theory, finite-simulation,
migration, and response uncertainties are retained without improvement.
In particular, the ATLAS signal-response uncertainty is not included in
the luminosity rescaling of the measured cross-section uncertainty.
The same signal model, event-level EFT truncation, and narrow-width
requirement are used for the Run~2 results and the HL-LHC projections.\\
The projected CMS exclusions in the left panel of
Fig.~\ref{fig:hllhcconstraints} extend down to
\begin{equation}
c_{t\gamma}\simeq
\left(
2.5\times10^{-7},\,
5.9\times10^{-7},\,
1.4\times10^{-6},\,
3.1\times10^{-6}
\right)~\GeV^{-1}
\label{eq:cmshlctgamma}
\end{equation}
at fixed $c_{tg}=1.0\times10^{-5}~\GeV^{-1}$ for
$m_T=500$, $750$, $1000$, and $1250~\GeV$, respectively.
At fixed
$c_{t\gamma}=1.0\times10^{-5}~\GeV^{-1}$, the projected exclusions
extend up to
\begin{equation}
c_{tg}\simeq
\left(
5.8\times10^{-4},\,
1.8\times10^{-4},\,
7.3\times10^{-5},\,
3.2\times10^{-5}
\right)~\GeV^{-1}.
\label{eq:cmshlctg}
\end{equation}

For the ATLAS projection in the right panel of
Fig.~\ref{fig:hllhcconstraints}, the projected exclusions at fixed
$c_{tg}=1.0\times10^{-5}~\GeV^{-1}$ extend down to
\begin{equation}
c_{t\gamma}\simeq
\left(
1.8\times10^{-6},\,
3.3\times10^{-6},\,
5.9\times10^{-6},\,
1.2\times10^{-5}
\right)~\GeV^{-1},
\label{eq:atlashlctgamma}
\end{equation}
while those at fixed
$c_{t\gamma}=1.0\times10^{-5}~\GeV^{-1}$ extend up to
\begin{equation}
c_{tg}\simeq
\left(
5.4\times10^{-5},\,
3.0\times10^{-5},\,
1.7\times10^{-5},\,
8.1\times10^{-6}
\right)~\GeV^{-1}.
\label{eq:atlashlctg}
\end{equation}
All quoted projected exclusion reaches are restricted to the
narrow-width domain.\\
The increased luminosity extends the sensitivity toward smaller values of
$c_{t\gamma}$ and, depending on the chosen slice, toward larger values of
$c_{tg}$. This asymmetric behavior is expected because the two coefficients
do not enter the fiducial rates in equivalent ways. The production
contributions, the radiative branching fractions, and the relative weights of
the CMS decay topologies all depend nonlinearly on the two coefficients.
Moreover, changing either coefficient changes $\Lambda_{\rm eff}$ and hence
the fraction of the EFT-dependent contribution that survives the event-level
truncation.\\
The projected improvement is therefore not a uniform rescaling of the
Run~2 contours. The CMS projection combines luminosity-scaled statistical
uncertainties with experimental systematic uncertainties reduced by a
factor of two, whereas the ATLAS projection rescales the total reported
experimental uncertainty. In both analyses, the unchanged SM and
signal-modeling uncertainties limit the gain. These projections represent
specified uncertainty scenarios rather than complete forecasts of future
detector performance, reconstruction efficiency, or theoretical precision.

\subsection{Interpretation in terms of physical decay fractions}
\label{sec:branchingresults}

The coupling-plane contours constitute the primary results because
$c_{t\gamma}$ and $c_{tg}$ enter directly in the production and decay
matrix elements. A branching-fraction representation is nevertheless useful
for comparison with searches expressed in terms of heavy-quark decay
probabilities.
For a compact and uniform comparison, we consider the nonzero benchmark
slice $c_{tg}=1.0\times10^{-5}~\GeV^{-1}$
and report the smallest $c_{t\gamma}$ value reached by the exclusion
for each analysis and mass hypothesis. The corresponding electromagnetic
branching fraction is evaluated at that contour point. This nonzero
benchmark avoids the special limiting behavior on the coordinate axes and
permits a consistent comparison between the CMS and ATLAS
reinterpretations. The results are summarized in
Table~\ref{tab:branchingresults}.

\begin{table}[ht!]
\centering
\caption{Observed Run~2 95\% CL exclusion reaches and projected expected
HL-LHC sensitivities in $c_{t\gamma}$ for the benchmark slice
$c_{tg}=1.0\times10^{-5}~\GeV^{-1}$, together with the corresponding
physical branching fraction $\mathcal B(T\to t\gamma)$. Each coupling
entry gives the smallest value reached by the exclusion along the
specified slice. Results are shown for the CMS $t\bar t\gamma$ and
ATLAS $t\bar t\gamma\gamma$ reinterpretations. The coupling values are
given in units of $10^{-5}~\GeV^{-1}$, and the branching fractions are
expressed as percentages. Values outside parentheses are the observed
Run~2 reaches, while values in parentheses are the projected expected
HL-LHC sensitivities. Only contour points satisfying the event-level
EFT-validity requirements and $\Gamma_T/m_T<0.1$ are reported.}
\label{tab:branchingresults}
\begin{tabular}{clcc}
\toprule
$m_T$ [GeV] &
Analysis &
$c_{t\gamma}$ exclusion reach
[$10^{-5}~\GeV^{-1}$] &
$\mathcal B(T\to t\gamma)$ [\%] \\
\midrule

500 &
CMS $t\bar t\gamma$ &
$0.047\;(0.025)$ &
$0.16\;(0.047)$ \\

500 &
ATLAS $t\bar t\gamma\gamma$ &
$0.23\;(0.18)$ &
$3.9\;(2.4)$ \\

\midrule

1000 &
CMS $t\bar t\gamma$ &
$0.21\;(0.14)$ &
$3.2\;(1.4)$ \\

1000 &
ATLAS $t\bar t\gamma\gamma$ &
$0.85\;(0.59)$ &
$35\;(21)$ \\

\bottomrule
\end{tabular}
\end{table}

\section{Conclusions}
\label{sec:conclusions}

We have reinterpreted the CMS $t\bar t\gamma$ differential measurement
and the ATLAS $t\bar t\gamma\gamma$ fiducial cross section as probes of
the electromagnetic and chromomagnetic transition interactions of a
charge-$2/3$ vector-like top partner, $T$. The analysis is formulated
within a restricted broken-phase phenomenological parametrization motivated
by the negligible-mixing limit. The conventional mixing-induced decays
$T\to Wb$, $T\to Zt$, and $T\to Ht$, together with the optional
charged-current transition in the more general model implementation, are
switched off. The resulting benchmark contains only the radiative decays
$T\to t\gamma$ and $T\to tg$.\\
The CMS reinterpretation uses the published absolute particle-level
$d\sigma/dp_T^\gamma$ spectrum and its bin-to-bin covariance matrix. The
heavy-$T$ prediction includes both the mixed
$T\bar T\to(t\gamma)(\bar t g)$ topology, together with its
charge-conjugate assignment, and the migration of double-radiative
$T\bar T\to t\bar t\gamma\gamma$ events into the
exactly-one-fiducial-photon category. These contributions are evaluated
independently using their topology-specific fiducial acceptances and physical
branching-fraction weights. The ATLAS result is treated as a fiducial
counting measurement, with a signal-specific, mass-dependent
detector-response transfer factor.\\
The higher-order QCD normalization is applied only to the pure-QCD
$T\bar T$ production component. Renormalization- and factorization-scale,
parton-distribution-function, finite-simulation, migration, and
reconstruction-response uncertainties are propagated consistently through
the statistical interpretation. The CMS and ATLAS constraints are reported
separately because the information required to determine the correlations
between the two measurements is not publicly available.\\
The validity of the effective description is assessed independently at every
coupling point. Contributions involving the nonrenormalizable production
interaction are truncated event by event by requiring
$M_{T\bar T}<\Lambda_{\rm eff}$. The physical interpretation
is further restricted to $\Gamma_T/m_T<0.1$, ensuring that $T$ remains a
well-defined resonance for which the factorized on-shell production-and-decay
treatment is expected to remain controlled. All nonzero coupling points
entering the reported constraints also correspond to prompt $T$ decays on
detector scales.\\
The observed Run~2 constraints and projected HL-LHC sensitivities are
presented as two-dimensional regions in the
$(c_{t\gamma},c_{tg})$ plane. For the representative slice
$c_{tg}=1.0\times10^{-5}~\GeV^{-1}$, the CMS Run~2 contours intersect
at approximately
$c_{t\gamma}=4.7\times10^{-7}~\GeV^{-1}$ at
$m_T=500~\GeV$ and
$2.1\times10^{-6}~\GeV^{-1}$ at $m_T=1000~\GeV$.
The corresponding ATLAS intersections are approximately
$2.3\times10^{-6}~\GeV^{-1}$ and
$8.5\times10^{-6}~\GeV^{-1}$, respectively. These contour points
correspond to electromagnetic branching fractions of approximately
$0.165\%$ and $3.26\%$ for CMS, and $3.86\%$ and $34.9\%$ for ATLAS.\\
For an integrated luminosity of $3000~\fb^{-1}$, scaling the CMS
statistical uncertainty with luminosity and halving its Run~2 experimental
systematic uncertainties, the projected CMS intersections at
$m_T=500$ and $1000~\GeV$ are approximately
$2.5\times10^{-7}$ and $1.4\times10^{-6}~\GeV^{-1}$,
respectively. With the total reported ATLAS experimental uncertainty
scaled by $\sqrt{140/3000}$, the corresponding ATLAS intersections are
approximately $1.8\times10^{-6}$ and
$5.9\times10^{-6}~\GeV^{-1}$. These values translate into projected
sensitivities to $\mathcal B(T\to t\gamma)$ of approximately
$0.0469\%$ and $1.42\%$ for CMS, and $2.37\%$ and $20.8\%$ for ATLAS.
These projections follow the uncertainty-scaling assumptions specified in
Sec.~\ref{sec:hllhcresults} and are not complete forecasts of future
experimental performance.\\
For both reinterpretations, statistically excluded regions satisfying the
EFT and narrow-width requirements are obtained through
$m_T=1250~\GeV$. For $m_T\geq1500~\GeV$, no valid exclusion remains.
At these higher masses, the steeply falling $T\bar T$ production rate
requires larger transition coefficients to produce an observable signal,
whereas larger coefficients lower $\Lambda_{\rm eff}$ and remove an
increasing fraction of the energetic EFT-dependent contribution. Smaller
coefficients improve the validity of the effective description but do not
produce sufficient sensitivity. The absence of a valid high-mass exclusion
therefore reflects the lack of overlap between the statistically excluded
region and the self-consistent domain of the effective calculation.\\
The two measurements probe complementary radiative configurations. The CMS
photon spectrum is sensitive primarily to the mixed photon--gluon topology,
with an additional contribution from double-radiative migration, whereas the
ATLAS measurement directly probes the double-photon topology. Their
reinterpretation demonstrates that existing top-associated photon
measurements provide meaningful sensitivity to nonstandard radiative
vector-like-top-partner interactions and complement dedicated searches
targeting conventional VLQ decay modes. A translation of these results to a
complete electroweak-gauge-invariant ultraviolet theory would require
matching the transition coefficients and consistently including any
additional interactions and decay channels predicted by that theory.

\section*{Data and code availability}
The experimental data used in this work are publicly available in the CMS
$t\bar t\gamma$ and ATLAS $t\bar t\gamma\gamma$ publications
\cite{CMS:2022lmh,ATLAS:2025aps} and their associated HEPData records.
The analysis codes and simulation configuration files are available from
the corresponding author upon reasonable request.

\section*{Acknowledgments}
We thank the authors of Ref.~\cite{Tong:2023lms} for making their model implementation available for cross-checks. 
The authors acknowledge the use of AI tools for language polishing.

\end{document}